\documentclass[]{aa}  

\usepackage{graphicx}
\usepackage[varg]{txfonts}
\usepackage[colorlinks=true,linkcolor=red,citecolor=blue,urlcolor=magenta]{hyperref}
\newcommand{\rev}[1]{#1}
\newcommand{\revrev}[1]{#1}
\newcommand{\revrevrev}[1]{#1}

\newcommand{\kms}{km\,s$^{-1}$}
\newcommand{\logg}{log\,$g$}
\newcommand{\loglxlbol}{log(L$_X$/L$_{bol}$)}

\newcommand{\masyr}{mas\,yr$^{-1}$}
\newcommand{\mjup}{$M_\mathrm{Jup}$ }

\newcommand{\vr}{$v_\mathrm{rad}$}
\newcommand{\pmdec}{$\mu_{\delta}$}
\newcommand{\pmra}{$\mu_{\alpha*}$}
\newcommand{\Teff}{$T_{\rm eff}$}
\newcommand{\teff}{$T_{\rm eff}$}

\bibpunct{(}{)}{;}{a}{}{,} 

\begin{document}

\title{Discovery of a directly imaged planet to the young solar analog \revrev{YSES~2}\thanks{Based on observations collected at the European Organisation for Astronomical Research in the Southern Hemisphere under ESO programs 099.C-0698(A), 0101.C-0153(A), 0101.C-0341(A), and 106.20X2.001.}}

\author{
Alexander~J.~Bohn\inst{1}
\and Christian~Ginski\inst{2,1}
\and Matthew~A.~Kenworthy\inst{1}
\and Eric~E.~Mamajek\inst{3,4}
\and Mark~J.~Pecaut\inst{5}
\and Markus~Mugrauer\inst{6}
\and Nikolaus~Vogt\inst{7}
\and Christian~Adam\inst{7,8}
\and Tiffany~Meshkat\inst{9}
\and Maddalena~Reggiani\inst{10}
\and Frans~Snik\inst{1}
}

\institute{Leiden Observatory, Leiden University, PO Box 9513, 2300 RA Leiden, The Netherlands\\
\email{bohn@strw.leidenuniv.nl}
\and Sterrenkundig Instituut Anton Pannekoek, Science Park 904, 1098 XH Amsterdam, The Netherlands
\and Jet Propulsion Laboratory, California Institute of Technology, 4800 Oak Grove Drive, M/S 321-100, Pasadena CA 91109, USA
\and Department of Physics \& Astronomy, University of Rochester, Rochester NY 14627, USA
\and Rockhurst University, Department of Physics, 1100 Rockhurst Road, Kansas City MO 64110, USA
\and Astrophysikalisches Institut und Universitäts-Sternwarte Jena, Schillergäßchen 2, D-07745 Jena, Germany
\and Instituto de Física y Astronomía, Facultad de Ciencias, Universidad de Valparaíso, Av. Gran Bretaña 1111, Playa Ancha, Valparaíso, Chile
\and Núcleo Milenio Formación Planetaria - NPF, Universidad de Valparaíso, Av. Gran Bretaña 1111, Valparaíso, Chile
\and IPAC, California Institute of Technology, M/C 100-22, 1200 East California Boulevard, Pasadena CA 91125, USA
\and Institute of Astronomy, KU Leuven, Celestijnenlaan 200D, B-3001 Leuven, Belgium
}

\date{Received February 6, 2021 / Accepted <date>}

\abstract 
{
To understand the origin and formation pathway of wide-orbit gas giant planets, it is necessary to expand the limited sample of these objects.
The mass of exoplanets derived with spectrophotometry, however, varies strongly as a function of the age of the system and the mass of the primary star.
} 
{
By selecting stars with similar ages and masses, the Young Suns Exoplanet Survey (YSES) aims to detect and characterize planetary-mass companions to solar-type host stars in the Scorpius-Centaurus association.
} 
{
\revrevrev{Our survey} is carried out with VLT/SPHERE with short exposure sequences on the order of 5\,min per star per filter.
The subtraction of the stellar point spread function (PSF) is based on reference star differential imaging (RDI) using the other targets (with similar colors and magnitudes) in the survey in combination with principal component analysis.
Two astrometric epochs that are separated by more than one year are used to confirm co-moving companions by proper motion analysis.
} 
{
We report the discovery of \revrev{YSES~2b}, a co-moving, planetary-mass companion to the K1 star \revrev{YSES~2} (TYC~8984-2245-1, 2MASS~J11275535-6626046).
The primary has a Gaia EDR3 distance of 110\,pc, and we derive a revised mass of $1.1\,M_\sun$ and an age of approximately 14\,Myr.
We detect the companion in two observing epochs southwest of the star at a position angle of 205$\degr$ and with a separation of $\sim1\farcs05$, which translates to a minimum physical separation of 115\,au at the distance of the system.
Photometric measurements in the $H$ and $K_s$ bands are indicative of a late $L$ spectral type, similar to the innermost planets around HR~8799.
We derive a photometric planet mass of $6.3^{+1.6}_{-0.9}\,M_\mathrm{Jup}$ using AMES-COND and AMES-dusty evolutionary models; 
this mass corresponds to a mass ratio of $q=(0.5\pm0.1)$\,\% with the primary.
This is the lowest mass ratio of a direct imaging planet around a solar-type star to date.
We discuss potential formation mechanisms and find that the current position of the planet is compatible with formation by disk gravitational instability, \rev{but its mass is lower than expected from numerical simulations}.
\rev{%
Formation via core accretion must have occurred closer to the star, yet we do not find evidence that supports the required outward migration, such as via scattering off another undiscovered companion in the system.
}
We can exclude \rev{additional} companions with masses \rev{greater} than 13\,\mjup in the full field of view of the detector ($0\farcs15<\rho<5\farcs50$), \rev{at 0\farcs5 we can rule out further objects that are more massive than 6\,\mjup}, and for projected separations $\rho>2\arcsec$ we are sensitive to planets with masses as low as 2\,\mjup.
} 
{
\revrev{YSES~2b} is an ideal target for follow-up observations to further the understanding of the physical and chemical formation mechanisms of  wide-orbit Jovian planets.
The YSES strategy of short snapshot observations ($\leq5$\,min) and PSF subtraction based on a large reference library proves to be extremely efficient and should be considered for future direct imaging surveys.
}


\maketitle

\section{Introduction}
\label{sec:introduction}

Despite several remarkable exoplanet \rev{and brown dwarf} discoveries by high-contrast imaging at high angular resolution in the past few years \citep[e.g.,][]{marois2008,Schmidt2008,marois2010,lagrange2010,rameau2013,bailey2014,macintosh2015,chauvin2017a,keppler2018,haffert2019,janson2019,bohn2020a,bohn2020c}, there is \rev{an ongoing debate regarding} the formation mechanisms that create these \rev{super-}Jovian gas giants with semimajor axes greater than 10\,au.
It is unclear whether these companions have a star-like origin from a collapsing molecular cloud that is broken up into fragments, creating planetary-mass objects similar to a stellar binary \citep[][]{kroupa2001, chabrier2003}, or through formation in a circumstellar disk instead.
The classical bottom-up framework postulates formation via core accretion by coagulation of small dust grains into planetary embryos \citep[][]{pollack1996,alibert2005,dodsonrobinson2009,lambrechts2012}.
These evolve either via collisions or pebble accretion \citep[][]{johansen2010,ormel2010} into planetary cores that are massive enough to accrete a gaseous envelope and to open a gap in the disk \citep[][]{paardekooper2004}.
In the corresponding top-down scenario, planetary cores can be created by gravitational instabilities leading to the collapse of dense regions in the protoplanetary disk \citep[][]{boss1997,rafikov2005,durisen2007,kratter2010,boss2011,kratter2016}.

To study this question for the underlying planet formation mechanisms from a statistical point of view, several direct imaging surveys have been conducted  \citep[e.g.,][]{vigan2012,galicher2016,bowler2016,vigan2017}.
Synthetic planet populations that represent each of the potential formation channels can be compared to the observational results from surveys and place constraints on the efficiency of the corresponding formation pathway \citep[e.g.,][]{mordasini2009a,mordasini2009b,forgan2013,forgan2018}.
The two largest surveys were carried out with two of the most advanced adaptive-optics assisted, high-contrast imagers available:
the Spectro-Polarimetric High- contrast Exoplanet REsearch \citep[SPHERE;][]{beuzit2019} instrument at the 8.2\,m ESO/VLT and the Gemini Planet Imager \citep[GPI;][]{macintosh2014} at the 8.1\,m Gemini South telescope.
The preliminary statistical analysis of the first 300 stars from the Gemini PLanet Imager Exoplanet Survey \citep[GPIES;][]{nielsen2019} concludes that giant planets between 10\,au and 100\,au that have masses smaller than $13\,M_\mathrm{Jup}$ favorably form via core accretion mechanisms, whereas brown dwarf companions in the same separation range but with masses from $13\,M_\mathrm{Jup}$ to $80\,M_\mathrm{Jup}$ seem to be predominantly created by disk instabilities.
This finding is supported by the analysis of the first 150 stars observed within the scope of the SpHere INfrared survey for Exoplanets \citep[SHINE;][]{vigan2020}, which additionally hypothesizes that companions with masses between $1\,M_\mathrm{Jup}$ and $75\,M_\mathrm{Jup}$ are likely to originate from bottom-up formation scenarios around B and A type stars, whilst objects of the same mass around M-type stars are consistent with simulated populations from top-down mechanisms.
For the intermediate masses of F-, G-, and K-type stars, the observed detections can be explained by a combination of both formalisms.
A statistical meta-analysis on the distribution of wide-orbit companion eccentricities carried out by \citet{bowler2020} provides supporting evidence for two distinct formation channels shaping the populations of giant planets ($2\,M_\mathrm{Jup}<M<15\,M_\mathrm{Jup}$) and brown dwarfs ($15\,M_\mathrm{Jup}<M<75\,M_\mathrm{Jup}$).

Most of these statistical evaluations are affected by the small number of actual substellar companions that were detected in the preceding imaging surveys.
To expand the sample size for solar-mass host stars, we started the Young Suns Exoplanet Survey \citep[YSES;][Bohn et al. in prep.]{bohn2020a} that is observing a homogeneous sample of 70 $\sim$15\,Myr-old, K-type stars in the Lower Centaurus Crux (LCC) subgroup of the Scorpius-Centaurus association \citep[Sco-Cen;][]{dezeeuw1999}.
All stars have masses close to $1\,M_\sun$ and the proximity \citep[average parallactic distance $\langle D\rangle=114\pm17$;][]{GaiaEDR3} and youth of the LCC facilitate the direct imaging search of giant, self-luminous substellar companions around these stars.

In this article we report the detection of a new exoplanet that was discovered within the scope of our survey.
\revrev{%
As this is already the second planetary system discovered by YSES, we introduce a new stellar identifier that is based on our survey acronym.
The details of this new designation are described in \S\ref{sec:yses_acronym}.
}
In \S\ref{sec:observations_and_data_reduction} we describe our observations and data reduction methods.
We discuss previous observations on the host star and reassess its main parameters in \S\ref{sec:stellar_properties}.
The results of our high-contrast imaging observations are presented and analyzed in \S\ref{sec:results_analysis}.
We discuss potential formation mechanisms of this newly detected exoplanet in \S\ref{sec:discussion} and we present our conclusions in \S\ref{sec:conclusions}.

\section{\revrev{Nomenclature of YSES planets}}
\label{sec:yses_acronym}

\revrev{
Owing to the recent success of YSES, we decided to introduce a dedicated catalog that will be used for star-planet systems discovered within the scope of our survey.
The YSES acronym has been verified by the IAU Commission B2 Working Group on Designations and was added to the Simbad database \citep[][]{wenger2000}.\footnote{Database entry available at: \url{http://cds.u-strasbg.fr/cgi-bin/Dic-Simbad?/18721212}}
The nomenclature of planet hosts from our survey is YSES~NNN and planets that are associated with these stars will be named YSES~NNNa, accordingly.
Following these guidelines, we assigned the host star of the intriguing multi-planet system that was discovered around \object{TYC~8998-760-1} the new primary identifier \object{YSES~1} \citep[][]{bohn2020a,bohn2020c}.
The planets formerly known as \object{TYC~8998-760-1~b} and \object{TYC~8998-760-1~c}, will be named \object{YSES~1b} and \object{YSES~1c}, henceforth.
Further planetary systems discovered by our survey will receive designated YSES identifiers followed by ascending integer identifications (IDs).
Hence, the new companion discovered within the scope of this paper will be referred to as \object{YSES~2b}, orbiting its Sun-like host \object{YSES~2}.
}

\section{Observations and data reduction} 
\label{sec:observations_and_data_reduction}

We observed \revrev{YSES~2} (\object{TYC~8984-2245-1}, \object{2MASS J11275535-6626046}) as part of YSES on the nights of 2018 April 30 (PI: Kenworthy) and 2020 December 8 (PI: Vogt) with SPHERE (mounted at the Nasmyth platform of Unit Telescope 3 of the ESO Very Large Telescope).
We used the IRDIS camera \citep[][]{dohlen2008} in classical imaging mode, applying a broadband filter in the $H$ and $K_s$ bands during the first and second nights, respectively.
The observations were carried out in pupil stabilized imaging mode and an apodized Lyot coronagraph was used to block the flux of the primary star \citep[][]{soummer2005,martinez2009,carbillet2011}.
In addition to the science frames, we obtained center frames, with a sinusoidal pattern applied to the deformable mirror that creates a waffle pattern to locate the position of the star behind the coronagraphic mask; sky frames of an offset position with no adaptive optics (AO) correction and without any source in the field of view, to subtract the instrument and thermal background; and non-coronagraphic images of the star that are used for photometric reference of point sources detected in the science images.
\rev{%
This last category of non-coronagraphic flux images was obtained with an additional neutral density filter in the optical path to record an unsaturated stellar point spread function (PSF) in the linear readout regime of the detector.
This neutral density filter (filter ID: ND\_1.0) provided an attenuation of $7.9$ and $6.9$ across the $H$ and $K_s$ bandpasses, respectively.
}
A detailed description of the observing setup and the weather conditions can be found in Appendix~\ref{sec:obervation_setup}.

The data reduction was performed with \texttt{PynPoint} \citep[version 0.8.1;][]{stolker2019} and included basic processing steps such as dark and flat calibration, bad pixel cleaning, sky subtraction, and correction for the instrumental distortion along the vertical axis of the detector.
To remove the stellar halo that is affecting approximately the innermost 1\farcs2 around the coronagraph, we utilized an approach based on reference star differential imaging \citep[RDI;][]{smith1984} in combination with principal component analysis \citep[PCA;][]{pynpoint,soummer2012}.
As the parallactic rotation of our YSES observations is usually less than a few degrees, classical PSF subtraction schemes, such as angular differential imaging \citep[ADI;][]{marois2006}, perform much worse compared to this reference library approach.
This method of combined RDI plus PCA was already successfully employed to recover circumstellar disks in archival HST data \citep[e.g.,][]{choquet2014} and within the scope of our survey for the discovery of a transition disk around the YSES target \object{Wray~15-788} \citep[][]{bohn2019}.

Owing to the same location on sky, similar distances, and spectral types, all YSES targets exhibit very similar magnitudes in the red part of the optical spectrum (where the wavefront sensor of SPHERE is operating) and at the near-infrared wavelengths of our scientific observations.
This facilitates comparable AO corrections amongst all our YSES observations, and the resulting images compose an excellent reference library to perform RDI.
The reference targets that were used for our library PSF subtraction are listed in Appendix~\ref{sec:reference_library}.
We modeled the stellar PSF with 50 principal components that were obtained from our full reference library.
After the PSF subtraction, the frames were de-rotated according to their parallactic angles and median combined.

For the astrometric calibration we used the standard instrumental solution as presented by \citet{maire2016} with a wavelength independent true north offset of $-1\fdg75\pm0\fdg08$ and plate scales of $(12.251\pm0.010)\,\mathrm{mas}\,\mathrm{px}^{-1}$ and $(12.265\pm0.010)\,\mathrm{mas}\,\mathrm{px}^{-1}$ in the $H$ and $K_s$ band, respectively.

\section{Stellar properties}   
\label{sec:stellar_properties}

We briefly summarize previous literature characterizing \revrev{YSES~2}
in \S\ref{subsec:prev} and compile the stellar properties of \revrev{YSES~2}
in Table \ref{tbl:stellar_properties}. 
In \S\ref{subsec:spectral_analysis_primary} and \S\ref{subsec:updated_stellar_parameters}, we derive updated stellar parameters for \revrev{YSES~2}, more importantly, including stellar mass and age.

\subsection{Previous studies\label{subsec:prev}} 

\revrev{YSES~2} was first identified as a young star in the Search for Associations Containing Young stars (SACY) survey \citep{Torres06} of optical counterparts to the ROSAT All-Sky Survey (RASS)
X-ray sources \citep{Voges99}.
\citet{Torres06} reported the star to be a Li-rich (EW[Li\,I $\lambda$6707] = 367 m\AA) K1V(e) star with filled-in H$\alpha$, showing fast rotation ($v\sin i$ = 19.3 km\,s$^{-1}$) and radial velocity 15.8\,$\pm$\,1.0 km\,s$^{-1}$.
Based on its position, proper motion, and youth indicators,  \citet{Preibisch08} included the star in a list of new members of the LCC subgroup of the Sco-Cen OB association (their Table 4), and provided initial estimates of isochronal age (16 Myr), mass ($1.1\,M_{\sun}$), and fractional X-ray luminosity (log($L_X$/$L_{bol}$) = -3.2).
\citet{Preibisch08} also predicted a kinematic distance of 109\,pc (based on the proper motion and space velocity of LCC), which compares remarkably well to the Gaia EDR3 parallactic distance \citep[$\varpi$ = 9.1537\,$\pm$\,0.0118 mas, $D$ = 109.25\,$\pm$\,0.14 pc;][]{GaiaEDR3}.
\citet{Kiraga12} reported the star to be a variable in the All Sky Automatic Survey (ASAS; IDed as ASAS J112755-6625.9) showing high amplitude (0.093 mag in V) and rapid rotation ($P_\mathrm{rot}$ = 2.7325\,d), which is consistent with the observed saturated X-ray emission \citep[and right near the median rotation period for Sco-Cen pre-main-sequence stars of $\langle P_\mathrm{rot} \rangle$ $\simeq$ 2.4\,d;][]{Mellon17}. 
\citet{Pecaut16} include the star in their age analysis of pre-main-sequence K stars across Sco-Cen, estimating an age of 23 Myr and a mass of $1.0\,M_{\sun}$. 
The star has subsequently appeared in multiple LCC membership lists  \citep{Gagne18, Goldman18, Damiani19}.
The status of this star as a pre-main-sequence member of LCC \rev{is strongly corroborated by a 99.9\,\% membership probability from the BANYAN $\Sigma$ algorithm \citep[][]{Gagne18} applied to the available \textit{Gaia} astrometry and radial velocities \citep[][]{gaia2018,GaiaEDR3}}.  

\begin{table}[hbt]
\caption{Stellar properties of \revrev{YSES~2}.}
\label{tbl:stellar_properties}
\def\arraystretch{1.2}
\setlength{\tabcolsep}{12pt}
\begin{tabular}{@{}lll@{}}
\hline
Parameter & Value & Ref.\\ 
\hline
Main identifier & \revrev{YSES~2} & (1)\\
TYCHO ID & TYC~8984-2245-1 & (2)\\
2MASS ID & J11275535-6626046 & (3)\\
GAIA EDR3 ID & 5236792880333011968 & (4)\\
$\alpha$ (J2000) [hh mm ss.sss] & 11 27 55.355 & (4)\\
$\delta$ (J2000) [dd mm ss.ss] & -66 26 04.50 & (4)\\
Spectral Type & K1V(e) & (5)\\
$\varpi$ [mas] & 9.1537\,$\pm$\,0.0118 & (4)\\
$D$ [pc] & 109.25\,$\pm$\,0.14 & (7)\\
\pmra\, [\masyr] & -34.025\,$\pm$\,0.013 & (4)\\
\pmdec\, [\masyr] & 2.319\,$\pm$\,0.011 & (4)\\
\vr\, [\kms] & 13.41\,$\pm$\,0.17 & (5)\\
$B$ [mag] & 11.819\,$\pm$\,0.010 & (8)\\
$V$ [mag] & 10.860\,$\pm$\,0.017 & (8)\\
$G$ [mag] & 10.525\,$\pm$\,0.003 & (4)\\
$I$ [mag] & 9.773\,$\pm$\,0.044 & (9)\\
$J$ [mag] & 9.006\,$\pm$\,0.026 & (3)\\
$H$ [mag] & 8.484\,$\pm$\,0.029  & (3)\\
$K_\text{s}$ [mag] & 8.358\,$\pm$\,0.029 & (3)\\
$W1$ [mag] & 8.323\,$\pm$\,0.014 & (10)\\
$W2$ [mag] & 8.351\,$\pm$\,0.008 & (10)\\
$W3$ [mag] & 8.258\,$\pm$\,0.019 & (11)\\
$W4$ [mag] & 7.929\,$\pm$\,0.118 & (11)\\
$P_{rot}$ [day] & 2.7325 & (9)\\
$v\sin i$ [\kms] & 19.3\,$\pm$\,0.5 & (5)\\
\loglxlbol\;[dex] & -3.07\,$\pm$\,0.23 & (9)\\
EW(Li\,I\,$\lambda$6707) [m\AA] & 364\,$\pm$\,0.05 & (6)\\
EW(H$\alpha$) [m\AA] & 0.0 & (5)\\
$U$ [\kms] & -10.10\,$\pm$\,0.08 & (7)\\
$V$ [\kms] & -18.93\,$\pm$\,0.12 & (7)\\
$W$ [\kms] & -5.60\,$\pm$\,0.09 & (7)\\
$A_V$ [mag] & 0.06$^{+0.03}_{-0.04}$ & (6)\\
$T_\mathrm{eff}$ [K] & 4749\,$\pm$\,40 & (6)\\
$m_\mathrm{bol}$ [mag] & 10.396\,$\pm$\,0.015 & (6)\\
$M_\mathrm{bol}$ [mag] & 5.204\,$\pm$\,0.016 & (6)\\
$\log\left(L/L_\mathrm{\sun}\right)$ [dex] & -0.1854\,$\pm$\,0.0063 & (6)\\
$R$ [$R_{\sun}$] & 1.193\,$\pm$\,0.022 & (6)\\
Mass [$M_\mathrm{\sun}$] & 1.10\,$\pm$\,0.03  & (6)\\
Age [Myr] & 13.9\,$\pm$\,2.3 & (6)\\
\hline
\end{tabular}
\tablebib{
(1)~This paper, see \S\ref{sec:yses_acronym}; 
(2)~\citet{Hog2000}; 
(3)~\citet{Cutri03}; 
(4)~\citet{GaiaEDR3}, and \textit{Gaia} EDR3 and DR2 ID \#s are the same; 
(5)~\citet{Torres06}; 
(6)~this paper, see \S\ref{subsec:updated_stellar_parameters}; 
(7)~distance and heliocentric Galactic Cartesian velocity calculated using Gaia EDR3 values ($D$ = 1/$\varpi$); 
(8)~\citet{Henden16}; 
(9)~\citet{Kiraga12};
(10)~\citet{Eisenhardt20};
(11)~\citet{Cutri14}.
}
\end{table}

\subsection{Spectral analysis of \revrev{YSES~2}}
\label{subsec:spectral_analysis_primary}

To check the previously published spectral properties of \revrev{YSES~2}, we examined two archival UVES spectra  from the ESO archive taken on UT 2007 May 2 (Program 079.C-0556(A); PI Torres).
The UVES spectra at resolution $R$ = 40,000 were convolved to lower resolution $R$ = 3,000 and compared to the grid of MK spectral standards from \citet{Pecaut16}. 
The blue spectrum, 3280\,\AA -- 4560\,\AA, is consistent with K0V, but the red spectrum, 4730\,\AA -- 6840\,\AA, appears to be K2V.  
The H$\alpha$ line exhibits marginal emission, similar to the filled-in emission reported by \citet{Torres06}.   
Hence, we infer a temperature type of K1$\pm$1 and confirm the type K1V(e) published by \citet{Torres06}.
From the original spectra, we independently measure the equivalent width of the Li\,I $\lambda$6707 feature to be 364\,$\pm$\,5 m\AA, by simultaneously fitting  Voigt profiles to the Li\,I feature and Fe\,I blend nearby \citep[see, e.g., ][]{soderblom1993}.
This is in good agreement with the 367\,m\AA\, reported by \citet{Torres06}. 
\rev{%
The assigned luminosity class that we adapted from \citet{Torres06} does not necessarily imply that \revrev{YSES~2} is on the main sequence rather than being a pre-main-sequence star.
Even though the luminosity and gravity indicators used by \citet{Torres06} were more in line with main-sequence dwarfs than subgiant or giant standard stars, more persuasive indicators such as the HRD position, Li absorption, X-ray emission, and the confirmed LCC membership clearly favor the pre-main-sequence evolutionary stage.
}

\subsection{Updated stellar parameters}
\label{subsec:updated_stellar_parameters}

\begin{figure}
\resizebox{\hsize}{!}{\includegraphics{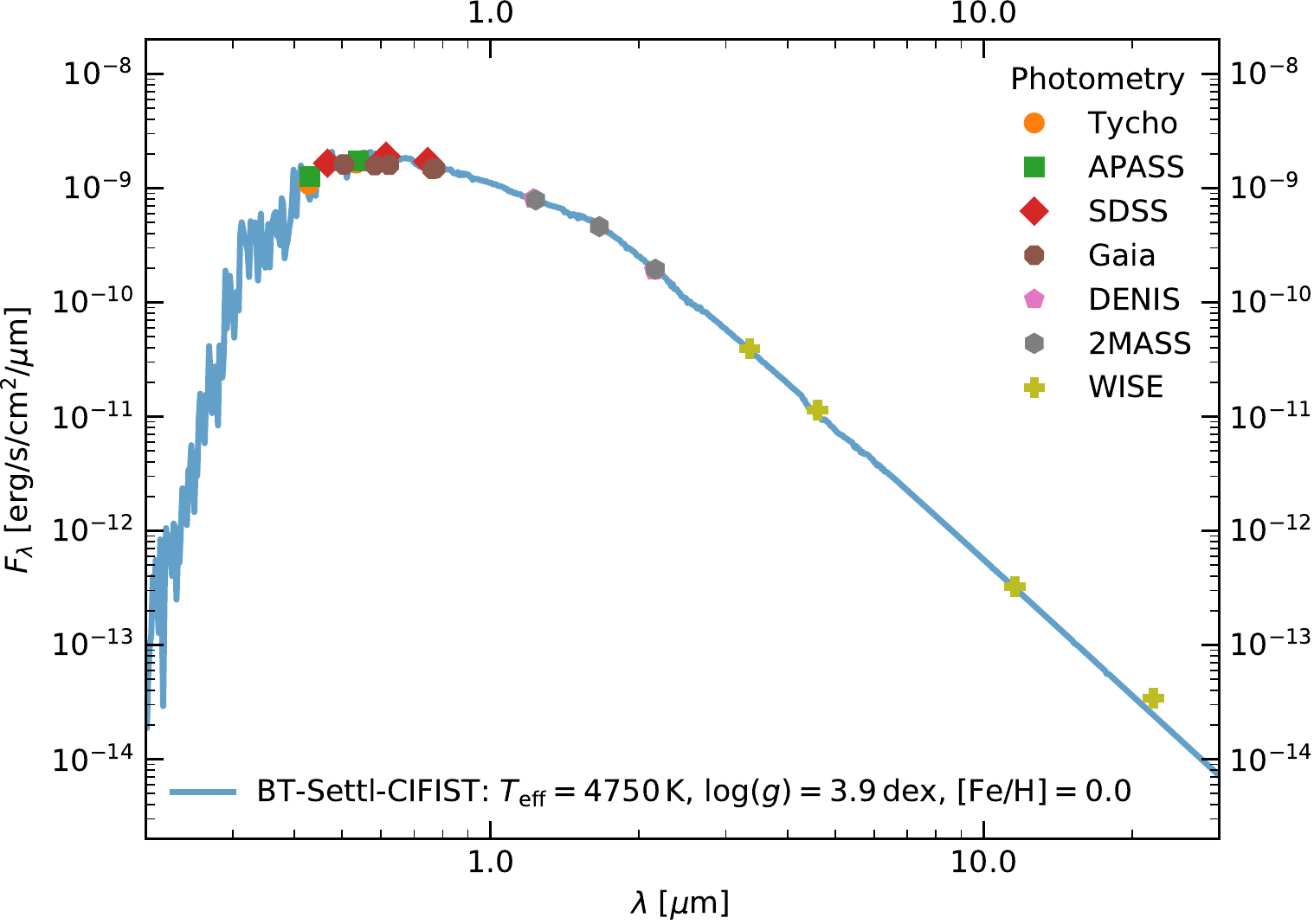}}
\caption{
Spectral energy distribution of \revrev{YSES~2}.
The colored markers indicate the archival photometric measurements of the star and the blue curve presents our best-fit BT-Settl-CIFIST model to the data.
The uncertainties of the photometric measurements are too small to be visualized in the figure.
}
\label{fig:tyc8984_sed}
\end{figure}

Using the VOSA spectral energy distribution (SED) analyzer \citep{Bayo08},\footnote{Online available at: \url{http://svo2.cab.inta-csic.es/theory/vosa/index.php}} we fit synthetic stellar spectra to the observed visible and infrared photometry for \revrev{YSES~2}.
For priors, we constrained the reddening to be E($B-V$) = 0.016\,$\pm$\,0.017 mag based on the STILISM 3D reddening maps \citep{Lallement19} and searched for best-fit synthetic spectra in the range 3000\,K $<$ \Teff\, $<$ 6000\,K, 3.5\,dex $<$ \logg\, $<$ 4.5\,dex, and metallicities -0.5 $<$ [Fe/H] $<$ 0.5 and [$\alpha$/Fe] = 0.0.
A Bayesian fit using the BT-Settl-CIFIST models using 22 photometric points yielded the following parameters: 
$A_V$ = 0.06 (0.02-0.09; 68\%CL; 0.00-0.11; 95\%CL), 
\teff\, = 4749\,K (4709-4789\,K; 68\%CL; 4700-4900\,K; 95\%CL),
\logg\, = 3.9\,dex (3.5-4.5\,dex), [Fe/H] = 0.0.
We present the results of this SED fit in Fig.~\ref{fig:tyc8984_sed}. 

The best-fit bolometric flux is $f_\mathrm{bol}$ = (1.7490\,$\pm$\,0.0248) $\times$10$^{-9}$ erg\,s$^{-1}$\,cm$^{-2}$, which on the IAU 2015 apparent bolometric magnitude scale translates to $m_\mathrm{bol}$ = 10.396\,$\pm$\,0.015\,mag.
Adopting the Gaia EDR3 parallax, this translates to absolute bolometric magnitude $M_\mathrm{Bol}$ = 5.204\,$\pm$\,0.016\,mag and bolometric luminosity log($L/L_{\odot}$) = -0.1854\,$\pm$\,0.0063. 
This is considerably more accurate than previous estimates (log($L/L_{\odot}$) = -0.06 \citep{Preibisch08}, log($L/L_{\odot}$) = -0.265\,$\pm$\,0.075 \citep{Pecaut16}) and benefits from a very precise distance, well-constrained extinction from 3D reddening maps, and integrating synthetic SEDs using 22 photometric data points. 
Combining this improved luminosity estimate with the improved \teff\, from the SED fitting (\teff\, = 4749\,$\pm$\,40\,K) yields a \rev{good} estimate of the radius of the star $(1.193\,\pm\,0.022\,R_{\sun}$).
A comparison against evolutionary tracks from \citet{baraffe2015} provided an updated stellar mass of $(1.10\pm0.03)\,M_{\sun}$ and an age of ($13.9\pm2.3$)\,Myr.
Furthermore we note that the SED of the star showed no signs of infrared excess through to the WISE-4 band (22\,$\mu$m).
The 2MASS-WISE colors
($K_s$-$W1$ = -0.035\,$\pm$\,0.032\,mag,
$K_s$-$W2$ = -0.007\,$\pm$\,0.030\,mag,
$K_s$-$W3$ = 0.100\,$\pm$\,0.035\,mag, 
$K_s$-$W4$ = 0.429\,$\pm$\,0.122\,mag) 
can be compared to the mean for K1 pre-main-sequence stars from \citet{Pecaut13} 
($K_s$-$W1$ = 0.09\,mag,
$K_s$-$W2$ = 0.06\,mag, 
$K_s$-$W3$ = 0.10\,mag, 
$K_s$-$W4$ = 0.18\,mag), and show a significant hint of IR excess. 
The $K_s$-$W4$ color is marginally red (2$\sigma$ excess), perhaps hinting at a debris disk (common among non-accreting pre-main-sequence Sco-Cen stars), but there is no corroborating evidence to further support this. 


We searched for common proper motion stellar or substellar companions for \revrev{YSES~2}. 
Assuming a mass of $1.0\,M_{\sun}$, the tidal radius of \revrev{YSES~2} is $\sim$1.35\,pc \citep[projected radius $\sim$0$^{\circ}$.71 or $\sim$2560\arcsec;][]{Mamajek13}, that is, bound companions would be expected to lie projected within this radius. 
Surveying the lists of Sco-Cen candidates and pre-main-sequence stars from the Gaia DR2 catalogs of \citet{Goldman18}, \citet{Zari18}, and \citet{Damiani19} with parallactic distances of $<$140 pc shows only a couple of likely LCC siblings within 1$^{\circ}$ of \revrev{YSES~2}: the poorly studied classical T Tauri star \object{Wray~15-813} at 2845\arcsec\, ($D$ = 101\,pc) \citep{Pereira03} and uncharacterized candidate pre-main-sequence object \object{2MASS~J11375287-6631197} ($D$ = 104\,pc). 
A query of the recently released Gaia EDR3 catalog searching for co-moving, co-distant objects (with generous selection range of proper motions in $\alpha$ and $\delta$ within $\pm$5 \masyr, and parallax $\pm$2 mas of \revrev{YSES~2}) yields
zero candidate companions within 1$^{\circ}$. 
Thus far, \revrev{YSES~2} appears to be a stellar singleton. 

\section{Observational results and analysis}
\label{sec:results_analysis}

\begin{figure*}
\resizebox{\hsize}{!}{\includegraphics{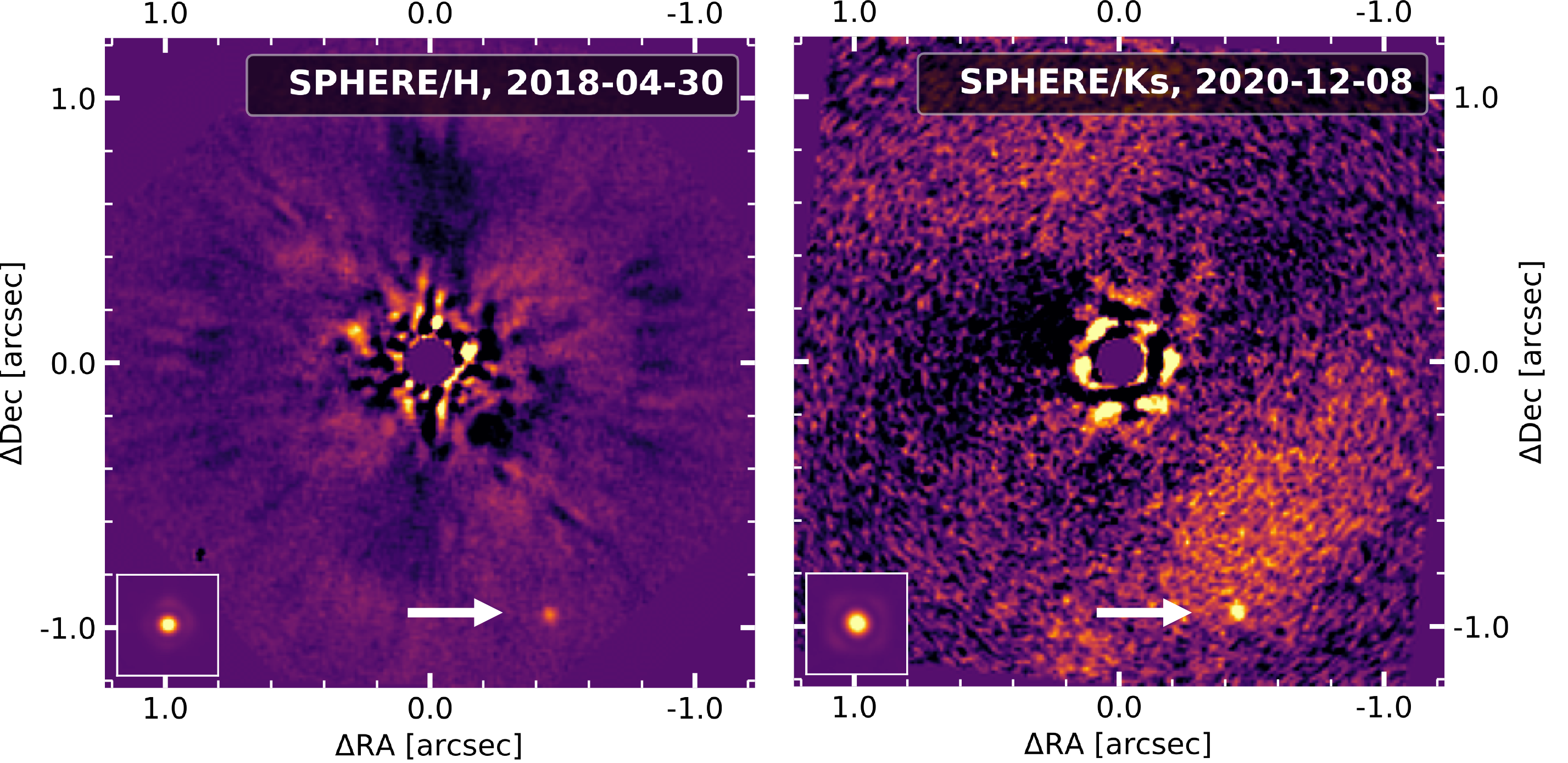}}
\caption{
Multi-epoch observations of \revrev{YSES~2} and its planetary-mass companion.
Final data products of the SPHERE observations collected in the $H$ band (left panel) and $K_s$ band (right panel) are presented.
For both filters, the stellar PSF is modeled by 50 principal components that were derived from a reference library of YSES targets.
These PSF models were subtracted and the residuals rotated such that north points up and east toward the left.
In the presented images the median of these de-rotated residuals is shown.
For the $K_s$ band data, uncorrected residuals of a wind-driven halo are detected that extend from the northeast to southwest.
The planet \revrev{YSES~2b} is highlighted by white arrows.
The primary is located at the origin of the coordinate system and we artificially masked the inner region up to the radial extent of the coronagraphic mask of 100\,mas.
\rev{%
To assess the spatial extent of the instrumental PSF, the median combination of the non-coronagraphic flux images of the primary star are shown in the lower left of each panel.
The intensity of each flux image is rescaled to match the maximum and minimum counts in the corresponding residual science image, and we display both images with the same spatial and color scales.
}
}
\label{fig:tyc8984_companion}
\end{figure*}

In \S\ref{subsec:results_analysis_astrometry} we show that our observations reveal a co-moving companion to \revrev{YSES~2}.
The reduced images for both epochs are presented in Fig.~\ref{fig:tyc8984_companion}.
Our photometric analysis in \S\ref{subsec:results_analysis_photometry} indicates that this companion has a mass that is significantly lower than the deuterium burning limit of $\sim13\,M_\mathrm{Jup}$.
We refer to this newly identified planet as \revrev{YSES~2b} henceforth.
In \S\ref{subsec:detection_limits} we present the detection and mass limits of our acquired data.

\subsection{Companion astrometry}
\label{subsec:results_analysis_astrometry}

\begin{figure}
\resizebox{\hsize}{!}{\includegraphics{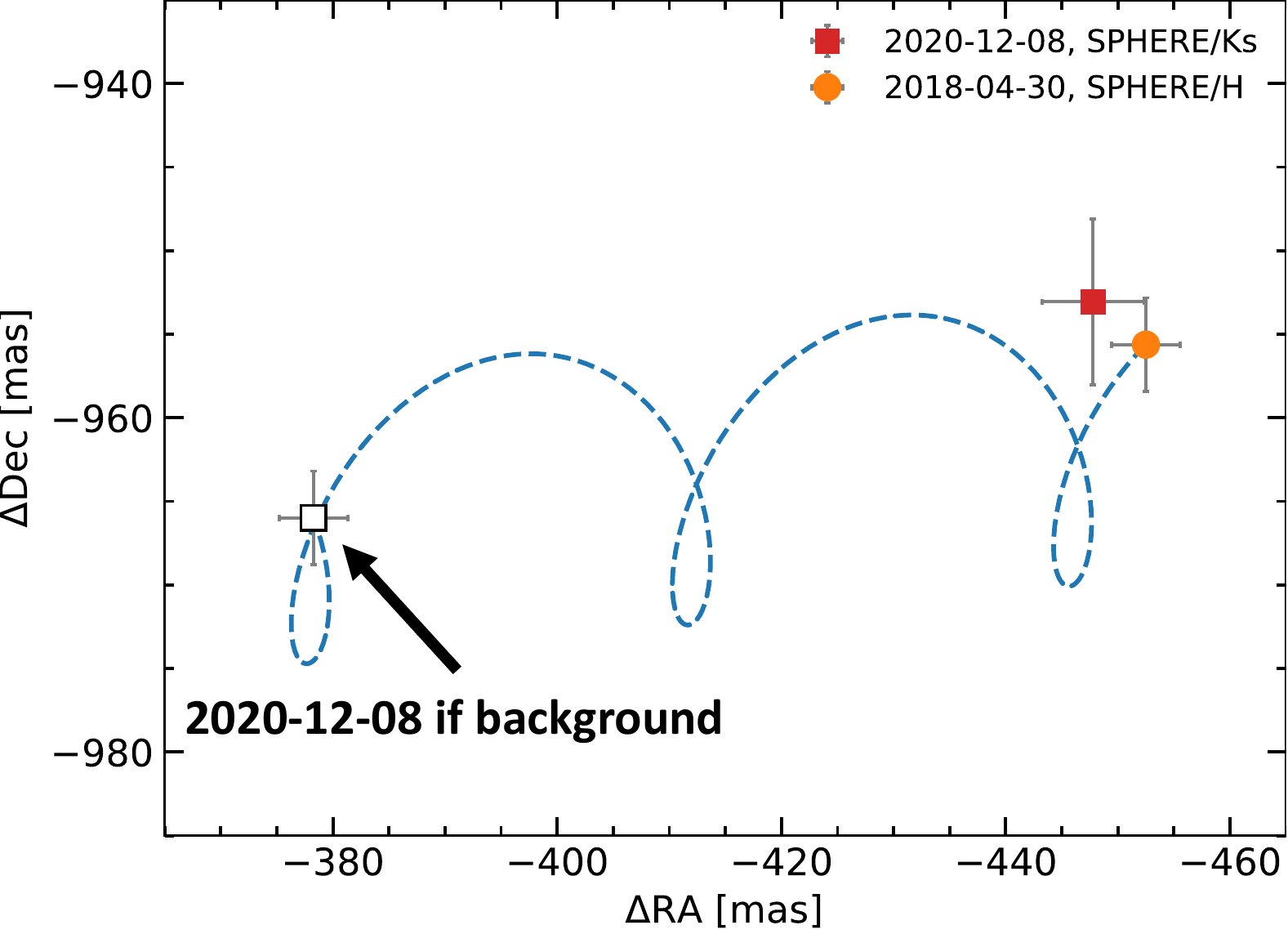}}
\caption{
Proper motion plot for \revrev{YSES~2b}.
The colored markers represent the relative astrometry with respect to the primary star measured for our two observational epochs.
The blue trajectory indicates the simulated motion of a static background object at infinity and the white marker is the theoretical position of such an object at our second observational epoch.
}
\label{fig:tyc8984_proper_motion}
\end{figure}

We extracted the companion astrometry and photometry by the injection of negative artificial companions \citep[e.g.,][]{lagrange2010,bonnefoy2011}.
A detailed description of our method is presented in Appendix~\ref{sec:extraction_astrometry_photometry}.
The extracted astrometry is listed in Table~\ref{tbl:companion_properties}.
\begin{table*}
\caption{
Astrometry, photometry, and derived masses of \revrev{YSES~2b}.
}
\label{tbl:companion_properties}
\def\arraystretch{1.2}
\setlength{\tabcolsep}{10pt}
\begin{tabular}{@{}llllllll@{}}
\hline\hline
Observation date & Filter & Separation & Position angle & $\Delta$Mag & $M_\mathrm{abs}$ & \multicolumn{2}{c}{Photometric mass}\\
& & & & & & COND & Dusty \\
(yyyy-mm-dd) & & (mas) & (\degr) & (mag) & (mag) & ($M_\mathrm{Jup}$) & ($M_\mathrm{Jup}$)\\
\hline
2018-04-30 & $H$ & $1057\pm3$ & $205.3\pm0.2$ & $10.37\pm0.11$ & $13.57\pm0.11$ &$5.3\pm0.5$ & $8.0\pm0.7$ \\
2020-12-08 & $K_s$ & $1053\pm5$ & $205.2\pm0.2$ & $9.55\pm0.11$ & $12.71\pm0.11$ &$6.1\pm0.6$ & $6.4\pm0.6$ \\
\hline
\end{tabular}
\end{table*}
As visualized in the proper motion plot in Fig.~\ref{fig:tyc8984_proper_motion} the companion is clearly incompatible with the calculated trajectory of a static background object at 14$\sigma$ significance.
The relative astrometric motion with respect to the primary is consistent with a comoving companion.
This conclusion is further confirmed by a similar analysis of other point sources within the detector field of view.
As presented in Appendix~\ref{sec:astrometry_background_objects}, all additional off-axis point sources are consistent with being non-moving background contaminants.
We thus conclude that \revrev{YSES~2b} is a gravitationally bound companion to its solar-mass host star.
From our astrometric measurements we derived a projected physical separation of approximately 115\,au.
Future astrometric measurements are required to derive meaningful constraints to the orbital parameters of this wide-orbit planet.

\subsection{Companion photometry}
\label{subsec:results_analysis_photometry}

\begin{figure}
\resizebox{\hsize}{!}{\includegraphics{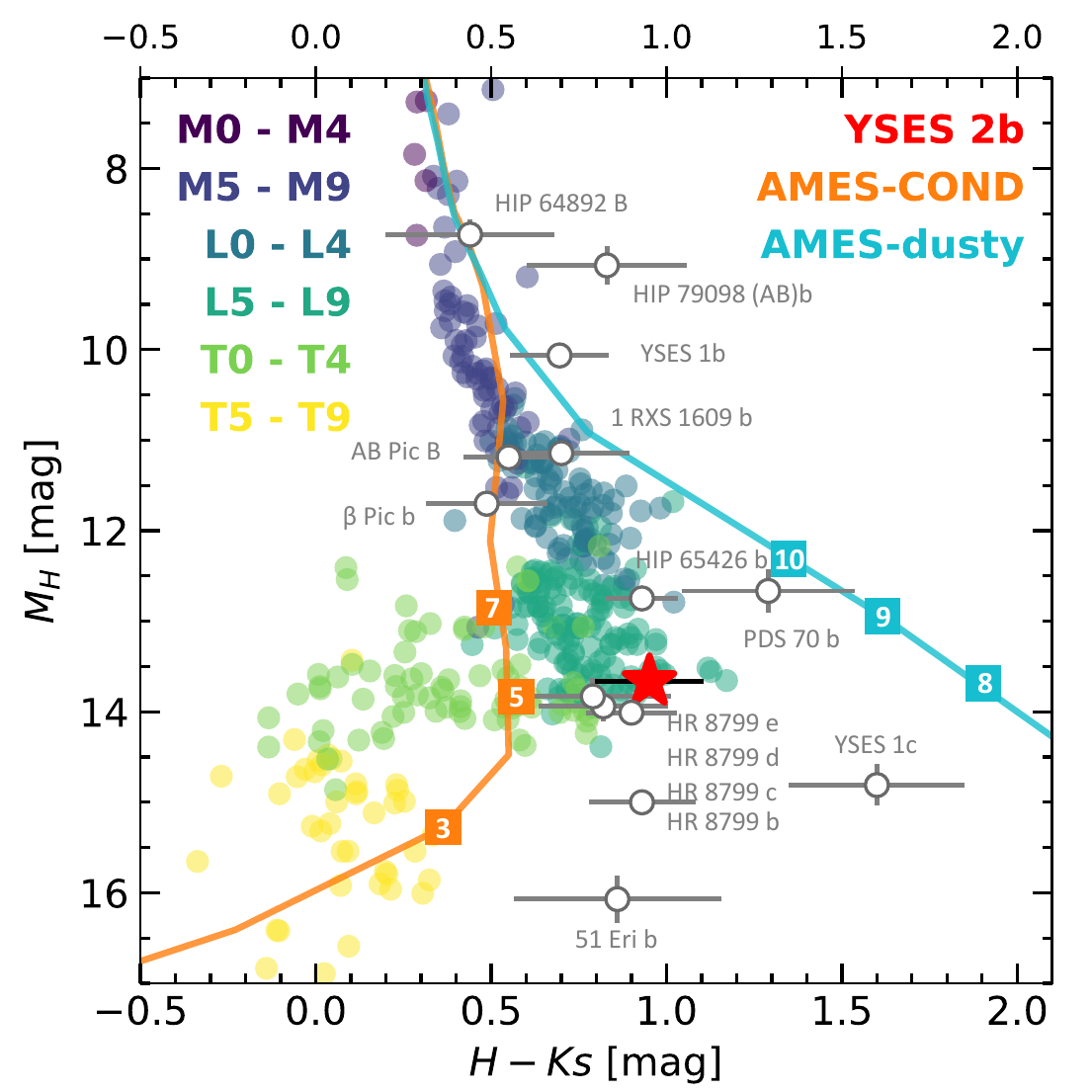}}
\caption{
Color-magnitude diagram for \revrev{YSES~2b}.
The colored markers show the sequence of field brown dwarfs with various spectral types from M to late T.
The white markers represent known directly imaged companions that are usually younger than the presented field objects.
\revrev{YSES~2b} is highlighted by the red star.
We further show AMES-COND and AMES-dusty evolutionary models that were evaluated at a system age of 13.9\,Myr (solid lines).
The markers along the line indicate the equivalent object masses in $M_\mathrm{Jup}$.
}
\label{fig:tyc8984_cmd}
\end{figure}

We present the photometry of the companion in Fig.~\ref{fig:tyc8984_cmd} in a color-magnitude diagram.
The corresponding numerical values are reported in Table~\ref{tbl:companion_properties}.
\revrev{YSES~2b} is consistent with a late L to early T spectral type when comparing it to colors of field brown dwarfs from the NIRSPEC Brown Dwarf Spectroscopic Survey \citep{mclean2003,mclean2007}, the IRTF spectral library \citep{rayner2009,cushing2005}, the L and T dwarf data archive \citet{knapp2004,golimowski2004,chiu2006}, and the SpeX Prism Libraries \citep{burgasser2010,gelino2010,burgasser2007,siegler2007,reid2006,kirkpatrick2006,cruz2004,burgasser2006,mcelwain2006,sheppard2009,looper2007,burgasser2008,looper2010,muench2007,dhital2011,kirkpatrick2010,burgasser2004}.
Object distances were derived from Gaia EDR3 \citep{GaiaEDR3}, the Brown Dwarf Kinematics Project \citep{faherty2009}, and the Pan-STARRS1 3$\pi$ Survey \citep{best2018}.
In color-magnitude space, \revrev{YSES~2b} is very close to the innermost three planets of the \object{HR~8799} multi-planetary system \citep[][]{marois2008, marois2010}.
These three planets are classified as mid to late $L$ type dwarfs \citep[e.g.,][]{greenbaum2018}, which agrees well with the sequence evolution of the adjacent field brown dwarfs from L to T spectral types.%
\footnote{%
\rev{%
We would like to note that the planets around HR~8799, although closely located to the sequence of field brown dwarfs in the selected SPHERE filters as presented in the color-magnitude diagram in Fig.~\ref{fig:tyc8984_cmd}, can have near-infrared colors in different passbands that are significantly distinct from those of their field dwarf analogs \citep[e.g.,][]{currie2011}.
}
}
A similar spectral type in this domain, therefore, seems very likely for \revrev{YSES~2b}, requiring confirmation by measurements at higher spectral resolution.
Whereas the masses of the spectrally similar trio of HR~8799 c, d, and e are in the range 7-12\,$M_\mathrm{Jup}$ \citep[][]{wang2018,marois2008, marois2010}, it is likely that \revrev{YSES~2b} has an even lower mass as the system age of $(13.9\pm2.3)$\,Myr is significantly younger than the age of HR~8799, which is claimed to be member of the Columba association with an age of 30--50\,Myr \citep[][]{zuckerman2011,bell2015}.
This is supported by the AMES-COND and AMES-dusty models \citep{allard2001,chabrier2000} that we present in Fig.~\ref{fig:tyc8984_cmd} for a system age of 13.9\,Myr.
An individual evaluation of these isochrones yielded masses from 5.3\,\mjup\;to 8.0\,\mjup\;as presented in Table~\ref{tbl:companion_properties}.
The uncertainties originate from the errors in the system age and planet magnitude that were propagated by a bootstrapping approach with 1,000 randomly drawn samples from Gaussian distributions around both parameters.
When combining the posterior distributions for the different models and filters we derived a final mass estimate of $6.3^{+1.6}_{-0.9}\,M_\mathrm{Jup}$ as the 68\% confidence interval around the median of the sample.
This estimate is based on broadband photometric measurements alone;
further spectral coverage of the planetary SED will be important to constrain its effective temperature, luminosity, surface gravity, and mass.

\subsection{Detection limits}
\label{subsec:detection_limits}

\begin{figure}
\resizebox{\hsize}{!}{\includegraphics{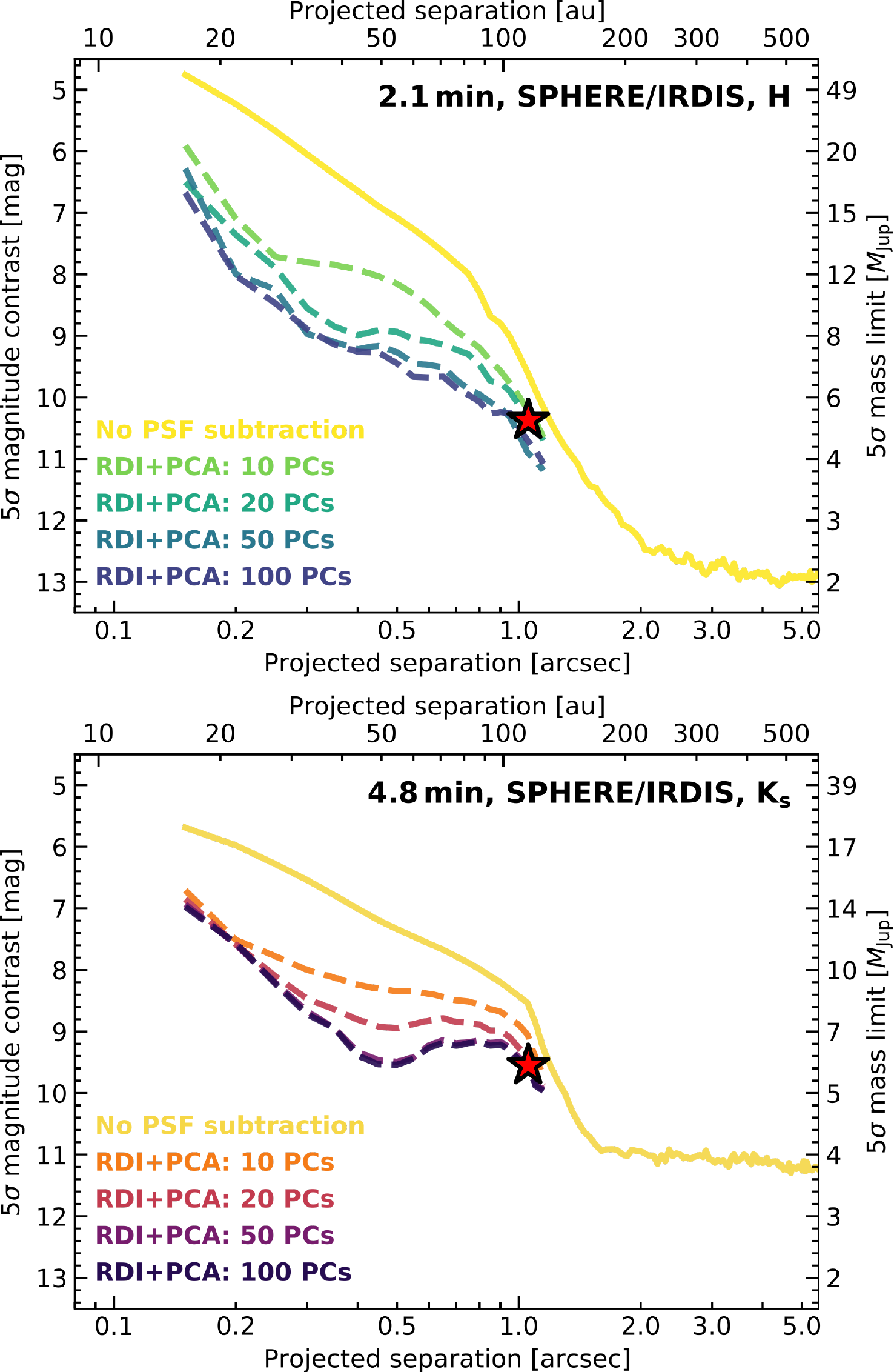}}
\caption{%
5$\sigma$ detection limits of our SPHERE/IRDIS observations in the $H$ (upper panel) and $K_s$ bands (lower panel).
On the left axis the magnitude contrast with respect to the primary star is reported, and the absolute magnitudes are converted to detectable planet masses with AMES-COND models as indicated on the right axis; this scale varies between the $H$ and $K_s$ bands.
The solid yellow lines represent the limits when no PSF subtraction is performed.
The dashed lines indicate the sensitivity, when a PSF subtraction with RDI plus PCA is performed.
The red star highlights the contrast of \revrev{YSES~2b} that we detect at $\sim5\sigma$ significance in both filters after PSF subtraction with more than 50 principal components.
}
\label{fig:detection_limits}
\end{figure}

To derive upper mass limits for additional companions in the system, we calculated the detection limits of our datasets.
As a baseline, we evaluated the contrast in the image that was obtained by de-rotating and median combining the individual exposures without any PSF subtraction.
This image covers the full field of view of the SPHERE/IRDIS detector up to an angular separation of 5\farcs5.
We evaluated the contrast directly in the final image using aperture photometry.
The chosen aperture size was one full width at half maximum (FWHM) of the unsaturated stellar PSF as measured in the flux images (see Table~\ref{tbl:observations} in Appendix~\ref{sec:obervation_setup}).
The signal flux was measured as the sum over the full circular aperture within the mean combined flux image and scaled for the flux difference with the science frames owing to the shorter exposure times and the applied neutral density filters.
For several radial positions that were equidistantly sampled from 0\farcs15 to 5\farcs50 in steps of 0\farcs05, we measured the noise as the standard deviation of the integrated flux within apertures that were distributed around the star at the same radial separation (excluding the signal aperture itself).
We applied a sigma clipping with an upper bound of 3$\sigma$ to the integrated fluxes of the noise apertures before calculating the standard deviation to discard apertures that were polluted by flux of off-axis point sources (see the full frame image in Fig.~\ref{fig:tyc8984_full} in Appendix~\ref{sec:astrometry_background_objects}).
A correction for small sample statistics as described in \citet{mawet2014} was considered in these noise calculations.
We reiterated this analysis for six uniformly spaced position angles and present the azimuthally averaged results as two-dimensional contrast curves in Fig.~\ref{fig:detection_limits}.
The solid yellow lines represent the \rev{5$\sigma$} raw contrast in the $H$ and $K_s$ bands that was obtained without any PSF subtraction.

For the innermost region around the star ($<1\farcs2$), the sensitivity was additionally assessed considering our PSF subtraction by RDI \revrevrev{plus} PCA. We used the \texttt{ContrastCurveModule} from \texttt{PynPoint} \rev{version 0.6.0}%
\footnote{\rev{%
As mentioned before, version 0.8.1 of \texttt{PynPoint} was used for all remaining analysis steps. 
The modules of both versions are compatible; only the implementation of some algorithms changed throughout the development process. 
This affects the \texttt{ContrastCurveModule}, which follows the iterative process described in this paragraph for release version 0.6.0.
We prefer this implementation over the solution presented in \texttt{PynPoint} version 0.8.1, which calculates one attenuation factor per position that is based on the single injection of an artificial companion with a user-defined signal-to-noise ratio.}} that utilizes the same aperture photometry framework and metric to evaluate the contrast for several positions that were distributed around the star in the residual images.
\rev{%
For each position, the module injects an artificial companion, whose detection significance is evaluated after the PSF subtraction with RDI combind with PCA.
In this framework, the signal aperture is directly placed on top of the position at which the artificial companion has been injected, and the noise apertures are azimuthally distributed around the primary star as described before, yet excluding the signal aperture itself.
The companion template was obtained as the median combination of the non-coronagraphic flux images that was scaled for the difference in exposure time and the attenuation due to the applied neutral density filter.
From an initial magnitude contrast of 8\,mag, the flux of the injected companion was adjusted and the post-processing was performed iteratively until the artificial companion was retrieved at 5$\sigma$ detection significance in the final image product.
These limiting magnitude contrasts were stored for each of the injection positions.
}
For the calculation of the final contrast curves with RDI \revrevrev{plus} PCA, we used a radial sampling in the range 0\farcs15-1\farcs20 with a spatial resolution of 0\farcs05, and six position angles that were equidistantly sampled in polar space.
\rev{%
Again, the contrast as a function of radial separation was obtained by azimuthal averaging of the various position angles.
}
We considered several numbers of principal components to model the stellar PSF as indicated by the sequentially colored, dashed lines in Fig.~\ref{fig:detection_limits}.
The detectable planet masses that correspond to the calculated magnitude contrasts were derived by evaluation of AMES-COND models at the system age of 13.9\,Myr (see right axes of Fig.~\ref{fig:detection_limits}).

The contrast performance close the star improves for an increasing number of principal components.
This differential gain in contrast ceases for $\sim$50 subtracted components and the contrast for 100 principal components does not change significantly compared to the curve generated for half as many components.
This justifies our previously selected value of 50 principal components that were used for our PSF subtraction with RDI \revrevrev{plus} PCA.
\rev{%
This amount of components is equivalent to 19\,\% and 30\,\% of the reference libraries in the $H$ and $K_s$ bands, which are composed of 269 and 164 individual frames, respectively}.
In the $H$ band we observe a contrast improvement of more than two magnitudes at an angular separation of 0\farcs2.
This corresponds to an increase in planet detection sensitivity by more than 45\,\mjup at this close separation.
The contrast improvement in the $H$ band is maximized at an angular separation of $\sim$0\farcs3, where RDI \revrevrev{plus} PCA provides detection limits that are approximately three magnitudes deeper than our raw data.
At angular separations larger than $1\arcsec$ the contrast improvement decreases as the flux contribution of the stellar PSF becomes negligible.
At separations $\geq2\arcsec$ we reach a fundamental noise floor that is mainly composed of residual sky background and detector read out noise.
The $K_s$ band contrast behaves very similar to the detection limits in the $H$ band and the RDI plus PCA reduction scheme can provide a maximum gain of up to 2.5\,mag at an angular separation of $\sim$0\farcs4.
The overall improvement for separations $<1\arcsec$ is marginally worse compared to the $H$ band data and the contribution of the asymmetric wind-driven halo is clearly visible for separations in the range $0\farcs5$-$1\farcs2$.
Combining the data from the $H$ and $K_s$ bands allows us to exclude \rev{stellar and brown dwarf} companions around \revrev{YSES~2} with masses $>13$\,\mjup for angular separations that are larger than 0\farcs15.
At 0\farcs5 we are sensitive to objects that are more massive than 6\,\mjup and for angular separations that are larger than $2\arcsec$ we can even rule out planets with masses as low as 2\,\mjup.

This demonstrates, impressively, how a large reference library can help to significantly improve the contrast performance at small angular separations $<1\farcs2$.
Especially for datasets with little parallactic rotation, RDI \revrevrev{plus} PCA should be considered as a default PSF subtraction strategy.
This conclusion is also supported by first results from the star-hopping mode that was recently implemented at VLT/SPHERE \citep[][]{wahhaj2021}.
As visualized by the red stars in Fig.~\ref{fig:detection_limits}, RDI \revrevrev{plus} PCA is required to detect \revrev{YSES~2b} at 5$\sigma$ significance in both the $H$ and $K_s$ band data.

\section{Discussion}
\label{sec:discussion}

The newly discovered planetary companion to \revrev{YSES~2} is among the lowest mass direct imaging companions known to date.
The only objects of similar low mass are 51\,Eri\,b \citep{macintosh2015}, HD\,95086\,b \citep{rameau2013}, HR8799\,b \citep{marois2008}, \rev{PDS~70\,b and c, \citep[][]{keppler2018,haffert2019,wang2020,stolker2020b},} and \revrev{YSES~1c} \citep[][]{bohn2020c}.
Of these, only \revrev{YSES~1c} is located around a solar-mass star.
Within the uncertainties, the mass of \revrev{YSES~2b} is the same as \revrev{YSES~1c}, which we previously discovered in our survey.
Given our mass estimate for the planet, the mass ratio of \revrev{YSES~2b} to its host star is $q=0.54^{+0.13}_{-0.08}$\,\%.
This value is comparable, but slightly lower than the $q=0.57\pm0.10$\,\% derived for \revrev{YSES~1c}.
The mass ratio is the lowest among direct imaging companions to solar-type stars.\footnote{See \citet{bohn2020a} for an overview of mass ratios of direct imaging companions to solar-type stars.}

The in situ formation of super-Jovian planets at tens or hundreds of \revrevrev{Astronomical Units} is challenging.
We recently discussed possible formation scenarios for such objects in the context of the \revrev{YSES~1} system in \cite{bohn2020a}.
In our previous study we considered scattering or planet capturing events to explain the current large separation of \revrev{YSES~1b}.
However, dynamic scattering by a third body in the system is expected to produce high eccentricities, inconsistent with orbital stability of both planetary components in the \revrev{YSES~1} system \citep[][]{bohn2020c}.\footnote{We, however, note that in some cases high eccentricities are inferred for substellar companions, for example, the brown dwarf companion to the young solar analog PZ\,Tel \citep[][]{Mugrauer2012,Ginski2014}.}
The new detection of \revrev{YSES~2b} in our small survey sample of 70 solar-type systems in Sco-Cen makes the hypothesis that we see captured free-floating planets unlikely.
\cite{goulinski2018} find with numerical simulations that only $\sim$0.1\% of solar-type stars in the Galactic thin disk should capture a free-floating planet in their lifetime.
\revrev{YSES~2b} in principle \rev{might have formed in situ via disk gravitational instability}.
\cite{boss2011} find that they can produce 1-5\,$M_\mathrm{Jup}$ planets between 30\,au and 70\,au with eccentricities as high as 0.35.
If \revrev{YSES~2b} is on an eccentric orbit it may explain its current projected separation of 110\,au. 
\cite{kratter2010} conversely find with their hydrodynamic simulations that planets formed via disk instability need to be at large separations outside of the 40\,au to 70\,au range to not accumulate too much mass and remain in the planetary regime.
\rev{%
This hypothesis that gravitationally instabilities predominantly create brown dwarf and stellar companions is supported by other theoretical studies \citep[e.g.,][]{zhu2012,forgan2013}.
}
Spatially resolved observations of gas-rich planet forming disks in the last few years have shown that radial substructures, which are thought to be caused by perturbing planets, are nearly ubiquitous (see, e.g., \citealt{Garufi2018} for an overview, observed in scattered light).
These structures can, in a growing number of cases, be traced out to tens or hundreds of \revrevrev{Astronomical Units} \citep[e.g.,][]{hltau2015,ginski2016a,deboer2016,Terwisga2018}.
Recently it was found that these substructures are already present in proto-stellar disks as young as $\sim$0.1\,Myr \citep{Sheehan2020a,Sheehan2020b}, suggesting that planet formation sets in early and operates on short timescales.
\rev{%
Despite this abundance of substructures observed in young, circumstellar environments, it is unlikely that all these protoplanetary disks are gravitationally unstable and support planet formation via this channel \citep[e.g.,][]{kratter2016}.
}
%
%
Conversely, we expect that the timescale to form a planet via core accretion at the current location of \revrev{YSES~2b} would be too long, given the system age and that the gas-rich disk in the system has already dissipated \citep[e.g.,][]{Haisch2001}.

\rev{%
Even though capturing scenarios are considered unlikely for \revrev{YSES~2b}, we cannot confidently conclude whether its formation via either top-down or bottom-up scenarios is more likely: whereas a mass of 6.3\,\mjup is rather low for an object to originate from gravitational disk instabilities, core accretion would favor a formation at closer separations to the star.
More data are thus required to explore the origin of this wide-orbit Jovian giant.
Promising methods to evaluate the likelihood of either formation scenario are by characterization measurements of the planetary atmosphere, continuous orbital monitoring to constrain especially its eccentricity, and deeper searches for additional companions in the system.
}

\rev{%
To identify the formation channel of \revrev{YSES~2b} via atmospheric characterization, we can utilize the framework postulated by \citet{oberg2011}, who argued that elemental abundances in the planetary atmosphere (and especially the C/O ratio) are directly linked to its natal environment in the planet-forming disk. 
Different ice lines in the protoplanetary disk and the associated freeze out of the corresponding molecular species alter molecular abundance ratios as a function of radial separation from the host star, making this atmospheric quantity a promising indicator of the natal environment and formation channel of a planet. 
The chemical and dynamical evolution of the disk can alter these initial abundances and should be considered in the analysis \citep[e.g.,][]{alidib2014,mordasini2016,eistrup2016,eistrup2018}. 
\citet{gravity_collaboration2020} utilized this framework to study $\beta$~Pic~b and proposed its formation through core accretion, with strong planetesimal enrichment based on its subsolar C/O abundance ratio.
}
If \revrev{YSES~2b} formed via disk gravitational instability, then we expect this object to have similar elemental abundances as the primary star in the system, while formation by core accretion should lead to an over-abundance of heavy elements due to pebble accretion.
\revrev{YSES~2b}, along with other planet mass objects detected by direct imaging, provides an ideal test case for future detailed atmospheric characterization.

\rev{%
We can further continue to monitor the separation and position angle of \revrev{YSES~2b} with respect to the primary star to derive orbital solutions for the planet \citep[e.g.,][]{wang2018}.
In particular, the VLTI/GRAVITY instrument \citep[][]{gravity_collaboration2017} will be extremely useful for this purpose, as it facilitates an unprecedented astrometric precision down to sub-milliarcsecond scales \citep[e.g.,][]{gravity_collaboration2019,wang2021}.
These astrometry measurements could be complemented by VLT/CRIRES$^+$ data to constrain the radial velocity of the planet and to obtain three-dimensional information about its orbital motion \citep[e.g.,][]{schwarz2016}.
The eccentricity of the planet might provide hints regarding the likelihood of a potential migration of the companion, which would be an indicator of formation via core accretion at closer separation to the star.
If this migration was caused by scattering off another, so far undetected companion to the primary star, a deep imaging campaign is required to search for evidence of such an additional component to the planetary system.
At the moment, we cannot provide conclusive evidence for the most likely formation scenario of \revrev{YSES~2b} based on the available data;
but future observations might be able to shed light on the origin of this Jovian gas giant.
}

\rev{%
Even though the second epoch observations of YSES are not concluded yet and candidate companions to $\sim$45 stars of our sample need to be confirmed or rejected, our survey has already discovered three planetary-mass companions amongst 70 young, Sun-like stars.
This high planet-detection rate is in stark contrast to previous surveys that were targeting Sun-like stars at closer distances than the LCC \citep[e.g.,][]{kasper2007,biller2013,galicher2016}, which discovered mostly stellar and brown dwarf companions.
These preliminary statistical results from YSES tentatively indicate that despite the farther distance, Sco-Cen and especially LCC are more favorable than moving groups in the immediate solar neighborhood for the detection of young planets briefly after their formation.
Many of the moving groups that were targeted in these aforementioned surveys are significantly older than LCC ($15\pm3$\,Myr), such as the Tucana-Horologium moving group \citep[$45\pm4$\,Myr][]{bell2015}, the AB Dor moving group \citep[$149^{+51}_{-19}$\,Myr;][]{bell2015}, or the Hercules-Lyra association \citep[$257\pm46$\,Myr;][]{eisenbeiss2013}.
As a consequence of the decreasing luminosity of objects below the deuterium burning limit with increasing age \citep[e.g.,][]{burrows1997}, it is natural that the sensitivity to Jovian planets is worse around members of these associations compared to significantly younger host stars.
Yet some of these closer moving groups have ages comparable to that of LCC -- such as the TW Hya association ($10\pm3$\,Myr) or the $\beta$ Pic moving group \citep[$24\pm3$\,Myr;][]{bell2015} -- and should provide even better planet-detection sensitivities owing to their much closer distances.
However, before speculating about potential reasons for this tentative overabundance of planetary-mass companions to our YSES targets, it is necessary to finish the second epoch observations of the survey and to derive reliable occurrence rates of planetary-mass companions to Sun-like stars in Sco-Cen.
}

\section{Conclusions}
\label{sec:conclusions}

We report the detection of a new directly imaged planet to the solar-mass primary \revrev{YSES~2} that was discovered within the scope of YSES.
Reassessment of the stellar parameters provided an effective temperature of $T_\mathrm{eff}=(4749\pm40)$\,K,  a luminosity of $\log\left(L/L_\sun\right)=-0.1854\pm0.0063$, a mass of ($1.10\pm0.03$)\,$M_\sun$, and a system age of ($13.9\pm2.3$)\,Myr.

We detect \revrev{YSES~2b} in two consecutive epochs collected on 2018 April 30 and 2020 December 18 with VLT/SPHERE.
The companion has a projected separation of approximately 1\farcs05, which translates to a physical minimum distance of $\sim$115\,au with respect to the primary star.
Photometric measurements in the $H$ and $K_s$ bands constrain a planet mass of $6.3^{+1.6}_{-0.9}\,M_\mathrm{Jup}$ according to AMES-COND and AMES-dusty evolutionary models.
This mass estimate is supported by the position of the object in color-magnitude space, where it is located amongst the mid to late $L$ type field brown dwarfs and close to HR~8799 c, d, and e.
The slightly higher mass estimates of these exoplanets on the order of 7--12$\,M_\mathrm{Jup}$ are consistent with the older system age of HR~8799 of 30--50\,Myr.

\rev{%
The mass and separation of \revrev{YSES~2b} are inconsistent with planet populations for most in situ formation scenarios:
whereas disk instabilities predominantly create companions above the deuterium burning limit at a separation of 110\,au, core-accretion mechanisms are not efficient enough to form a planet of 6.3\,\mjup this widely separated from the primary star.
So, the new companion might be either at the low-mass end of potential in situ formation outcomes from top-down scenarios, or it formed via core accretion at closer separation to the star and migrated to its current location.
Atmospheric characterization measurements of molecular abundance ratios, orbital monitoring, and evaluation of the eccentricity of the planet, or a deep search for additional companions in the system, might help to evaluate the likelihood of these potential formation pathways.
}
While we cannot rule out scattering or capture scenarios, we point out that the former require an (as of yet) third undetected body in the system, while the latter are unlikely given numerical simulations.
\revrev{YSES~2b} is an important addition to the sparsely populated group of wide-orbit gas giant companions.
Owing to the moderate separation with respect to the primary star, spectroscopic observations with JWST, VLT/ERIS, or VLTI/GRAVITY will be easily available.
These data will be important to further constrain the properties of this Jovian companion.
Measurements of molecular abundance ratios such as C/O \rev{or its orbital eccentricity} might even facilitate hypotheses regarding the most likely formation mechanism for this wide-orbit gas giant planet.

Our data rule out \rev{brown dwarf and stellar} companions with $M>13$\,\mjup in the SPHERE/IRDIS field of view for angular separations $>0\farcs15$ \rev{and at 0\farcs5 we can exclude objects that are more massive than 6\,\mjup}.
At separations that are larger than $2\arcsec$ we are even sensitive to planets with masses as low as 2\,\mjup.
In general, the applied PSF subtraction scheme based on RDI plus PCA is extremely successful and provides substantial contrast improvements ($>1$\,mag) for separations that are smaller than 1\arcsec.
In the $H$ band, the PSF subtraction enhances our sensitivity by more than 45\,\mjup at 0\farcs2, and the greatest contrast improvement of $\sim$3\,mag is achieved at an angular separation of 0\farcs3.
Our YSES strategy with short snapshot observations of $\leq5$\,min combined with a large reference library for PSF subtraction is certainly a promising approach to image planetary-mass companions to young, Sun-like stars in Sco-Cen.
With three newly discovered planetary-mass companions in less than 40\,h of allocated telescope time the survey efficiency is unprecedented and the mission concept can certainly be applied to future high-contrast imaging studies targeting different samples of pre-main-sequence stars.

\begin{acknowledgements}
\revrev{
We would like to thank the anonymous referee for the very valuable feedback that helped improving the quality of the manuscript.
Especially, the extremely kind way of providing this feedback was highly appreciated by the authors.
}
The research of AJB and FS leading to these results has received funding from the European Research Council under ERC Starting Grant agreement 678194 (FALCONER).
Part of this research was carried out at the Jet Propulsion Laboratory, California Institute of Technology, under a contract with the National Aeronautics and Space Administration (80NM0018D0004).
MM would like to thank the German Research Foundation (DFG) for support in the program MU 2695/27-1.
CA acknowledges support by ANID -- Millennium Science Initiative Program -- NCN19\_171.
MR acknowledges support from the FWO research program under project 1280121N.
This research has used the SIMBAD database, operated at CDS, Strasbourg, France \citep{wenger2000}.
This work has used data from the European Space Agency (ESA) mission {\it Gaia} (\url{https://www.cosmos.esa.int/gaia}), processed by the {\it Gaia} Data Processing and Analysis Consortium (DPAC, \url{https://www.cosmos.esa.int/web/gaia/dpac/consortium}).
Funding for the DPAC has been provided by national institutions, in particular the institutions participating in the {\it Gaia} Multilateral Agreement.
This publication makes use of VOSA, developed under the Spanish Virtual Observatory project supported by the Spanish MINECO through grant AyA2017-84089.
To achieve the scientific results presented in this article we made use of the \emph{Python} programming language\footnote{Python Software Foundation, \url{https://www.python.org/}}, especially the \emph{SciPy} \citep{virtanen2020}, \emph{NumPy} \citep{numpy}, \emph{Matplotlib} \citep{Matplotlib}, \emph{emcee} \citep{foreman-mackey2013}, \emph{scikit-image} \citep{scikit-image}, \emph{scikit-learn} \citep{scikit-learn}, \emph{photutils} \citep{photutils}, and \emph{astropy} \citep{astropy_1,astropy_2} packages.

\end{acknowledgements}

\bibliographystyle{aa} 
\bibliography{mybib} 

\begin{appendix}

\section{Observational conditions and setup}
\label{sec:obervation_setup}

We present the observational setup and the weather conditions for our SPHERE observations in Table~\ref{tbl:observations}.
\begin{table*}
\caption{
SPHERE observations of \revrev{YSES~2}.
}
\label{tbl:observations}
\def\arraystretch{1.2}
\setlength{\tabcolsep}{10pt}
\begin{tabular}{@{}llllllll@{}}
\hline\hline
Observation date & Filter & FWHM\tablefootmark{a} & NEXP$\times$NDIT$\times$DIT\tablefootmark{b} & $\Delta\pi$\tablefootmark{c} & $\langle\omega\rangle$\tablefootmark{d} & $\langle X\rangle$\tablefootmark{e} & $\langle\tau_0\rangle$\tablefootmark{f} \\
(yyyy-mm-dd) & & (mas) & (1$\times$1$\times$s) & (\degr) & (\arcsec) & & (ms)\\
\hline
2018-04-30 & $H$ & 50.5 & 4$\times$1$\times$32 & 0.98 & 0.87 & 1.343 & 6.25\\
2020-12-08 & $K_s$ & 61.7 & 1$\times$18$\times$16 & 1.37 & 0.55 & 1.52 & 4.30 \\
\hline
\end{tabular}
\tablefoot{
\tablefoottext{a}{
Full width at half maximum measured for the non-coronagraphic stellar PSF.
}
\tablefoottext{b}{
NEXP describes the number of exposures, NDIT is the number of subintegrations per exposure, and DIT is the detector integration time of an individual subintegration.
}
\tablefoottext{c}{
$\Delta\pi$ describes the amount of field rotation during the observation, if it is carried out in pupil-stabilized mode (only valid for CI observations).
}
\tablefoottext{d}{$\langle X\rangle$ denotes the average airmass during the observation.}
\tablefoottext{e}{$\langle\omega\rangle$ denotes the average seeing conditions during the observation.}
\tablefoottext{f}{$\langle\tau_0\rangle$ denotes the average coherence time during the observation.}
}
\end{table*}

\section{Reference library}
\label{sec:reference_library}

The reference libraries were compiled from the full amount of YSES data that were collected under ESO IDs 099.C-0698(A) (PI: Kenworthy), 0101.C-0153(A) (PI: Kenworthy), 0101.C-0341(A) (PI: Bohn), and 106.20X2.001 (PI: Vogt).
We used RDI in the innermost region of the images $<1\farcs2$, where the stellar halo was dominating the received flux.
We deselected all targets with obvious point sources or extended structures in this region because these signals are not part of the stellar PSF and would therefore deteriorate the quality of our model created by PCA.
The remaining targets, their observation epochs, and observing conditions that were used as a reference library for the $H$ and $K_s$ band data are listed in Tables~\ref{tbl:reference_library_h} and \ref{tbl:reference_library_k}, respectively.
\rev{%
In the $H$ band we have 269 individual reference frames and in  the $K_s$ band we have 164.
}

\begin{table*}
\caption{
Reference library for the data reduction in $H$ band.
}
\label{tbl:reference_library_h}
\def\arraystretch{1.2}
\setlength{\tabcolsep}{10pt}
\scalebox{0.9}{%
\begin{tabular}{@{}llllll@{}}
\hline\hline
Target & Observation date & NEXP$\times$NDIT$\times$DIT\tablefootmark{a} & $\langle\omega\rangle$\tablefootmark{b} & $\langle X\rangle$\tablefootmark{c} & $\langle\tau_0\rangle$\tablefootmark{d} \\
(2MASS ID) & (yyyy-mm-dd) & (1$\times$1$\times$s) & (\arcsec) & & (ms)\\
\hline
J11272881-3952572 & 2017-04-18 & 4$\times$1$\times$32 & 1.51 & 1.10 & 1.40 \\
J11320835-5803199 & 2017-06-17 & 4$\times$1$\times$32 & 0.67 & 1.47 & 2.90 \\
J11445217-6438548 & 2018-05-14 & 4$\times$1$\times$32 & 0.72 & 1.31 & 2.38 \\
J11454278-5739285 & 2018-06-04 & 4$\times$1$\times$32 & 0.70 & 1.19 & 2.80 \\
J11454278-5739285 & 2019-01-13 & 4$\times$1$\times$32 & 1.14 & 1.62 & 3.83 \\
J12065276-5044463 & 2017-04-02 & 3$\times$1$\times$32 & 1.24 & 1.11 & 1.50 \\
J12090225-5120410 & 2018-05-15 & 4$\times$1$\times$32 & 0.86 & 1.12 & 2.70 \\
J12090225-5120410 & 2019-12-14 & 12$\times$2$\times$32 & 0.63 & 1.51 & 7.75 \\
J12101065-4855476 & 2017-04-18 & 4$\times$1$\times$32 & 1.71 & 1.15 & 1.40 \\
J12113142-5816533 & 2018-12-22 & 3$\times$2$\times$32 & 1.46 & 1.47 & 2.13 \\
J12113142-5816533 & 2019-02-18 & 4$\times$2$\times$32 & 0.45 & 1.23 & 14.30 \\
J12160114-5614068 & 2018-12-27 & 4$\times$2$\times$32 & 0.41 & 1.45 & 11.88 \\
J12164023-7007361 & 2018-12-23 & 3$\times$1$\times$32 & 0.98 & 1.59 & 2.93 \\
J12164023-7007361 & 2019-02-15 & 4$\times$1$\times$32 & 0.54 & 1.63 & 11.20 \\
J12185802-5737191 & 2017-06-17 & 2$\times$1$\times$32 & 0.72 & 1.22 & 2.70 \\
J12195938-5018404 & 2018-12-30 & 4$\times$1$\times$32 & 0.53 & 1.62 & 8.00 \\
J12210499-7116493 & 2019-01-12 & 4$\times$2$\times$32 & 0.80 & 1.53 & 4.25 \\
J12220430-4841248 & 2017-04-18 & 3$\times$1$\times$32 & 1.82 & 1.17 & 1.40 \\
J12234012-5616325 & 2017-06-17 & 4$\times$1$\times$32 & 0.62 & 1.72 & 3.45 \\
J12264842-5215070 & 2018-12-30 & 4$\times$1$\times$32 & 0.40 & 1.38 & 8.20 \\
J12302957-5222269 & 2018-12-30 & 4$\times$1$\times$32 & 0.38 & 1.33 & 9.85 \\
J12333381-5714066 & 2019-01-01 & 4$\times$1$\times$32 & 0.76 & 1.37 & 7.03 \\
J12333381-5714066 & 2019-01-14 & 4$\times$1$\times$32 & 1.26 & 1.21 & 2.45 \\
J12361767-5042421 & 2018-12-30 & 4$\times$1$\times$32 & 0.51 & 1.59 & 4.68 \\
J12361767-5042421 & 2019-12-18 & 16$\times$2$\times$32 & 1.14 & 1.57 & 3.07 \\
J12374883-5209463 & 2018-12-30 & 4$\times$1$\times$32 & 0.41 & 1.50 & 7.30 \\
J12383556-5916438 & 2019-01-03 & 4$\times$1$\times$32 & 0.52 & 1.59 & 13.90 \\
J12383556-5916438 & 2019-01-12 & 4$\times$1$\times$32 & 0.79 & 1.26 & 4.25 \\
J12393796-5731406 & 2017-06-17 & 4$\times$1$\times$32 & 0.64 & 1.77 & 3.83 \\
J12404664-5211046 & 2018-04-30 & 4$\times$1$\times$32 & 0.74 & 1.13 & 7.05 \\
J12442412-5855216 & 2017-06-17 & 4$\times$3$\times$32 & 0.71 & 1.37 & 2.67 \\
J12454884-5410583 & 2018-04-30 & 4$\times$1$\times$32 & 0.71 & 1.15 & 6.92 \\
J12480778-4439167 & 2017-06-17 & 4$\times$2$\times$32 & 0.90 & 1.34 & 2.75 \\
J12505143-5156353 & 2019-01-12 & 4$\times$1$\times$32 & 1.14 & 1.32 & 3.75 \\
J12510556-5253121 & 2019-01-08 & 4$\times$1$\times$32 & 0.58 & 1.68 & 3.90 \\
J13015069-5304581 & 2019-01-08 & 4$\times$1$\times$32 & 0.55 & 1.60 & 3.95 \\
J13055087-5304181 & 2018-06-11 & 4$\times$1$\times$32 & 0.82 & 1.14 & 1.95 \\
J13055087-5304181 & 2018-07-04 & 4$\times$1$\times$32 & 1.73 & 1.14 & 1.70 \\
J13064012-5159386 & 2018-04-30 & 4$\times$1$\times$32 & 0.56 & 1.13 & 8.15 \\
J13065439-4541313 & 2018-04-08 & 4$\times$1$\times$32 & 0.46 & 1.09 & 5.65 \\
J13095880-4527388 & 2018-05-01 & 4$\times$1$\times$32 & 1.08 & 1.07 & 2.70 \\
J13103245-4817036 & 2018-05-01 & 4$\times$1$\times$32 & 1.03 & 1.09 & 3.30 \\
J13121764-5508258 & 2017-08-31 & 4$\times$1$\times$32 & 0.68 & 2.22 & 4.42 \\
J13121764-5508258 & 2018-05-15 & 4$\times$1$\times$32 & 0.62 & 1.16 & 2.50 \\
J13174687-4456534 & 2018-05-28 & 4$\times$1$\times$32 & 0.70 & 1.07 & 4.33 \\
J13334410-6359345 & 2017-07-05 & 4$\times$1$\times$32 & 1.06 & 1.53 & 3.05 \\
J13343188-4209305 & 2017-04-02 & 4$\times$1$\times$32 & 1.14 & 1.21 & 1.70 \\
J13354082-4818124 & 2017-04-02 & 4$\times$1$\times$32 & 1.06 & 1.30 & 2.08 \\
J13380596-4344564 & 2017-04-02 & 4$\times$1$\times$32 & 1.05 & 1.32 & 2.40 \\
J13455599-5222255 & 2018-04-28 & 4$\times$1$\times$32 & 0.64 & 1.13 & 6.35 \\
\hline
\end{tabular}}
\tablefoot{
\tablefoottext{a}{
NEXP describes the number of exposures, NDIT is the number of subintegrations per exposure, and DIT is the detector integration time of an individual subintegration.
}
\tablefoottext{b}{$\langle X\rangle$ denotes the average airmass during the observation.}
\tablefoottext{c}{$\langle\omega\rangle$ denotes the average seeing conditions during the observation.}
\tablefoottext{d}{$\langle\tau_0\rangle$ denotes the average coherence time during the observation.}
}
\end{table*}

\begin{table*}
\caption{
Reference library for the data reduction in $K_s$ band.
}
\label{tbl:reference_library_k}
\def\arraystretch{1.2}
\setlength{\tabcolsep}{10pt}
\begin{tabular}{@{}llllll@{}}
\hline\hline
Target & Observation date & NEXP$\times$NDIT$\times$DIT\tablefootmark{a} & $\langle\omega\rangle$\tablefootmark{b} & $\langle X\rangle$\tablefootmark{c} & $\langle\tau_0\rangle$\tablefootmark{d} \\
(2MASS ID) & (yyyy-mm-dd) & (1$\times$1$\times$s) & (\arcsec) & & (ms)\\
\hline
J11445217-6438548 & 2018-05-14 & 4$\times$1$\times$32 & 0.77 & 1.31 & 2.60 \\
J11454278-5739285 & 2019-01-13 & 4$\times$1$\times$32 & 1.18 & 1.59 & 3.58 \\
J12090225-5120410 & 2018-05-15 & 4$\times$1$\times$32 & 0.70 & 1.12 & 2.90 \\
J12113142-5816533 & 2018-12-22 & 4$\times$2$\times$32 & 1.38 & 1.44 & 2.05 \\
J12113142-5816533 & 2019-02-18 & 4$\times$2$\times$32 & 0.45 & 1.22 & 15.00 \\
J12160114-5614068 & 2018-12-27 & 4$\times$2$\times$32 & 0.47 & 1.42 & 10.27 \\
J12164023-7007361 & 2018-12-23 & 4$\times$1$\times$32 & 1.06 & 1.58 & 3.43 \\
J12164023-7007361 & 2019-02-15 & 4$\times$1$\times$32 & 0.57 & 1.61 & 10.75 \\
J12195938-5018404 & 2018-12-30 & 4$\times$1$\times$32 & 0.55 & 1.59 & 9.00 \\
J12210499-7116493 & 2019-01-12 & 4$\times$2$\times$32 & 0.82 & 1.52 & 4.40 \\
J12264842-5215070 & 2018-12-30 & 4$\times$1$\times$32 & 0.41 & 1.36 & 9.20 \\
J12302957-5222269 & 2018-12-30 & 4$\times$1$\times$32 & 0.45 & 1.32 & 7.48 \\
J12333381-5714066 & 2019-01-01 & 4$\times$1$\times$32 & 0.80 & 1.36 & 6.25 \\
J12333381-5714066 & 2019-01-14 & 4$\times$1$\times$32 & 1.24 & 1.21 & 2.30 \\
J12333381-5714066 & 2020-12-10 & 1$\times$20$\times$16 & 0.58 & 1.76 & 5.50 \\
J12361767-5042421 & 2018-12-30 & 4$\times$1$\times$32 & 0.47 & 1.56 & 6.22 \\
J12374883-5209463 & 2018-12-30 & 4$\times$1$\times$32 & 0.46 & 1.48 & 6.95 \\
J12383556-5916438 & 2019-01-03 & 4$\times$1$\times$32 & 0.46 & 1.56 & 12.47 \\
J12383556-5916438 & 2019-01-12 & 4$\times$1$\times$32 & 0.94 & 1.26 & 3.45 \\
J12404664-5211046 & 2018-04-30 & 4$\times$1$\times$32 & 0.87 & 1.13 & 7.10 \\
J12454884-5410583 & 2018-04-30 & 4$\times$1$\times$32 & 0.66 & 1.15 & 8.97 \\
J12505143-5156353 & 2019-01-12 & 4$\times$1$\times$32 & 1.03 & 1.31 & 4.10 \\
J12510556-5253121 & 2019-01-08 & 4$\times$1$\times$32 & 0.52 & 1.65 & 3.98 \\
J13015069-5304581 & 2019-01-08 & 4$\times$1$\times$32 & 0.49 & 1.58 & 4.80 \\
J13055087-5304181 & 2018-06-11 & 4$\times$1$\times$32 & 0.93 & 1.14 & 2.02 \\
J13055087-5304181 & 2018-07-04 & 4$\times$1$\times$32 & 1.73 & 1.14 & 1.70 \\
J13064012-5159386 & 2018-04-30 & 4$\times$1$\times$32 & 0.56 & 1.13 & 9.88 \\
J13065439-4541313 & 2018-04-08 & 4$\times$1$\times$32 & 0.55 & 1.09 & 4.68 \\
J13095880-4527388 & 2018-05-01 & 4$\times$1$\times$32 & 1.03 & 1.07 & 2.45 \\
J13103245-4817036 & 2018-05-01 & 4$\times$1$\times$32 & 0.87 & 1.10 & 4.40 \\
J13121764-5508258 & 2018-05-15 & 4$\times$1$\times$32 & 0.62 & 1.16 & 3.00 \\
J13174687-4456534 & 2018-05-28 & 4$\times$1$\times$32 & 0.67 & 1.07 & 4.15 \\
J13455599-5222255 & 2018-04-28 & 4$\times$1$\times$32 & 0.65 & 1.13 & 6.03 \\
\hline
\end{tabular}
\tablefoot{
\tablefoottext{a}{
NEXP describes the number of exposures, NDIT is the number of subintegrations per exposure, and DIT is the detector integration time of an individual subintegration.
}
\tablefoottext{b}{$\langle X\rangle$ denotes the average airmass during the observation.}
\tablefoottext{c}{$\langle\omega\rangle$ denotes the average seeing conditions during the observation.}
\tablefoottext{d}{$\langle\tau_0\rangle$ denotes the average coherence time during the observation.}
}
\end{table*}

\section{Extraction of companion astrometry and photometry}
\label{sec:extraction_astrometry_photometry}

We extracted the companion astrometry and photometry with the \texttt{SimplexMinimizationModule} of \texttt{PynPoint}.
This injects an artificial planet into the data prior to the stellar PSF subtraction with RDI plus PCA.
The planet template PSF is obtained from the unsaturated, non-coronagraphic flux images that were taken alongside the observations.
The methods injects the artificial planet into the data at the approximate position and magnitude of the real point source, considering the parallactic rotation during the observing sequence.
The PSF subtraction is performed using the same library as before (see Appendix~\ref{sec:reference_library}), we smooth the image with a Gaussian kernel with a FWHM of 12\,mas (which corresponds to the size of a detector pixel) to reduce pixel-to-pixel variations, and we evaluate the residuals in an aperture with a diameter of $\sim0\farcs25$ around the injection position.
We choose the image curvature, which is represented by the determinant of the Hessian matrix as function of merit, which we aim to minimize by varying the input separation, position angle, and magnitude contrast of our artificial companion.
We do not use the absolute value norm as presented by \citet{wertz2017} as an objective to the minimization because this would not consider large-scale features in the residual image that are not correctly modeled by our PSF subtraction approach.
Such a feature is for instance the asymmetric wind driven halo \citep[][]{cantalloube2018} that is apparent in the  $K_s$ band data in the northeastern to southwestern direction (see right panel of Fig.~\ref{fig:tyc8984_companion}).
This uncorrected stellar flux contributes to the planet signal and minimization of the absolute value norm around the planet position would certainly overestimate its flux and perhaps even compromise its astrometry.
Planet separation, position angle, and magnitude contrast are optimized simultaneously by a Nelder-Mead simplex minimization algorithm \citep[][]{nelder1965}.

Owing to the optimization process our final values for the planetary astrometry and photometry do not exhibit any intrinsic uncertainties.
To derive the systematic uncertainties of our injection and minimization approach, we follow the analysis described by \citet{stolker2020a}, using the cube in which the optimized negative planet is injected such that no companion signal remains in the data.
For 24 position angles that are equidistantly distributed in polar space we inject positive artificial companions into the data using the same magnitude contrast and the same radial separation as previously determined for our companion.
We extract the astrometry and photometry of these artificial companions with the same method as described before and we evaluate the deviations from the injection position and flux.
The standard deviation along the 24 distinct positions is utilized as uncertainty of our extraction method.
These are combined with additional astrometric uncertainties originating from the detector plate scale, the true north offset, and the centering accuracy of 2.5\,mas (see SPHERE manual) to derive the final value of planet separation and position angle as presented in Table~\ref{tbl:companion_properties}.
For the companion photometry, we add uncertainties due to the variation of the unsaturated stellar PSF throughout the sequence of flux measurements and we account for transmissivity variations of the neutral density filter across the broadband filter that was used for our observations (either the $H$ or $K_s$ band).

\section{Astrometric analysis of background objects}
\label{sec:astrometry_background_objects}

\begin{figure}
\resizebox{\hsize}{!}{\includegraphics{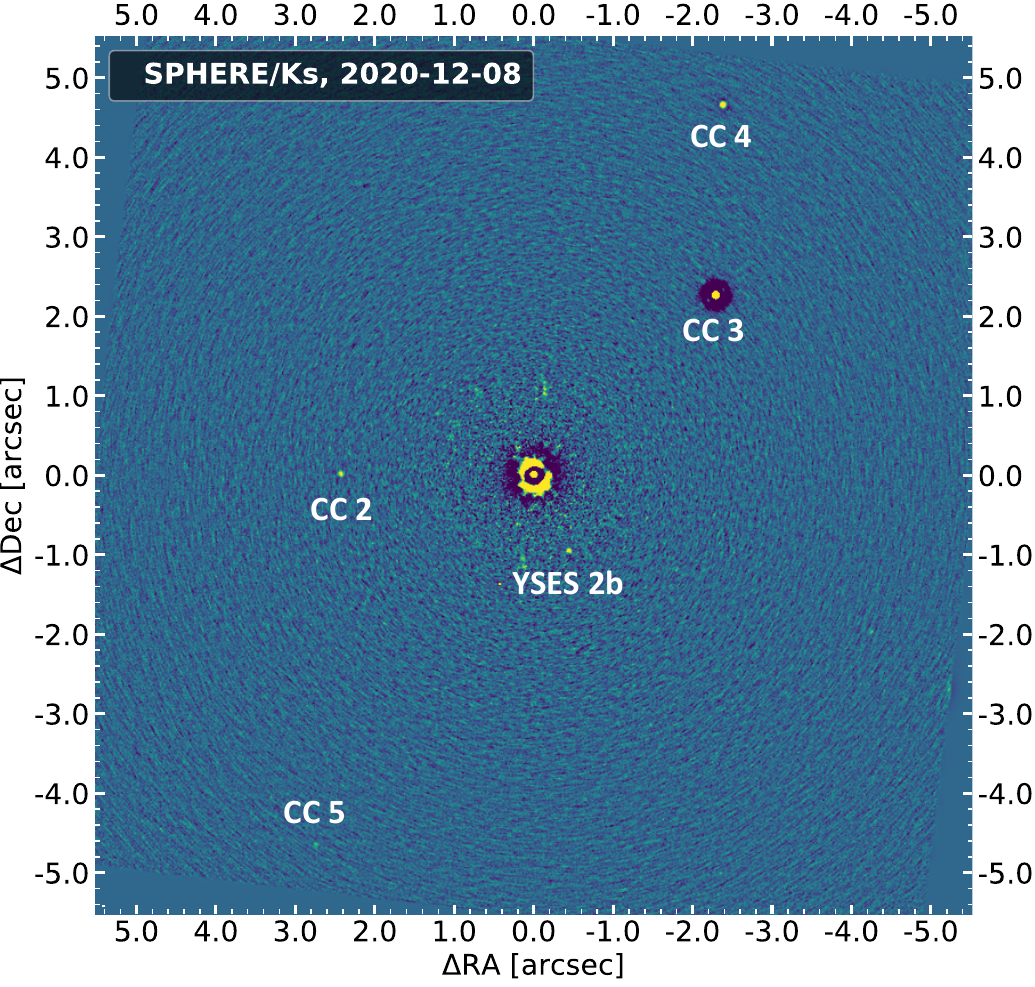}}
\caption{
Reduced SPHERE data for the full IRDIS field of view.
The images are de-rotated and median combined; an unsharp mask is applied to remove the stellar halo.
Four additional companion candidates to \revrev{YSES~2b} are identified in the field of view. 
The image is presented at an arbitrary logarithmic color scale to highlight the off-axis point sources.
}
\label{fig:tyc8984_full}
\end{figure}

\begin{figure}
\resizebox{\hsize}{!}{\includegraphics{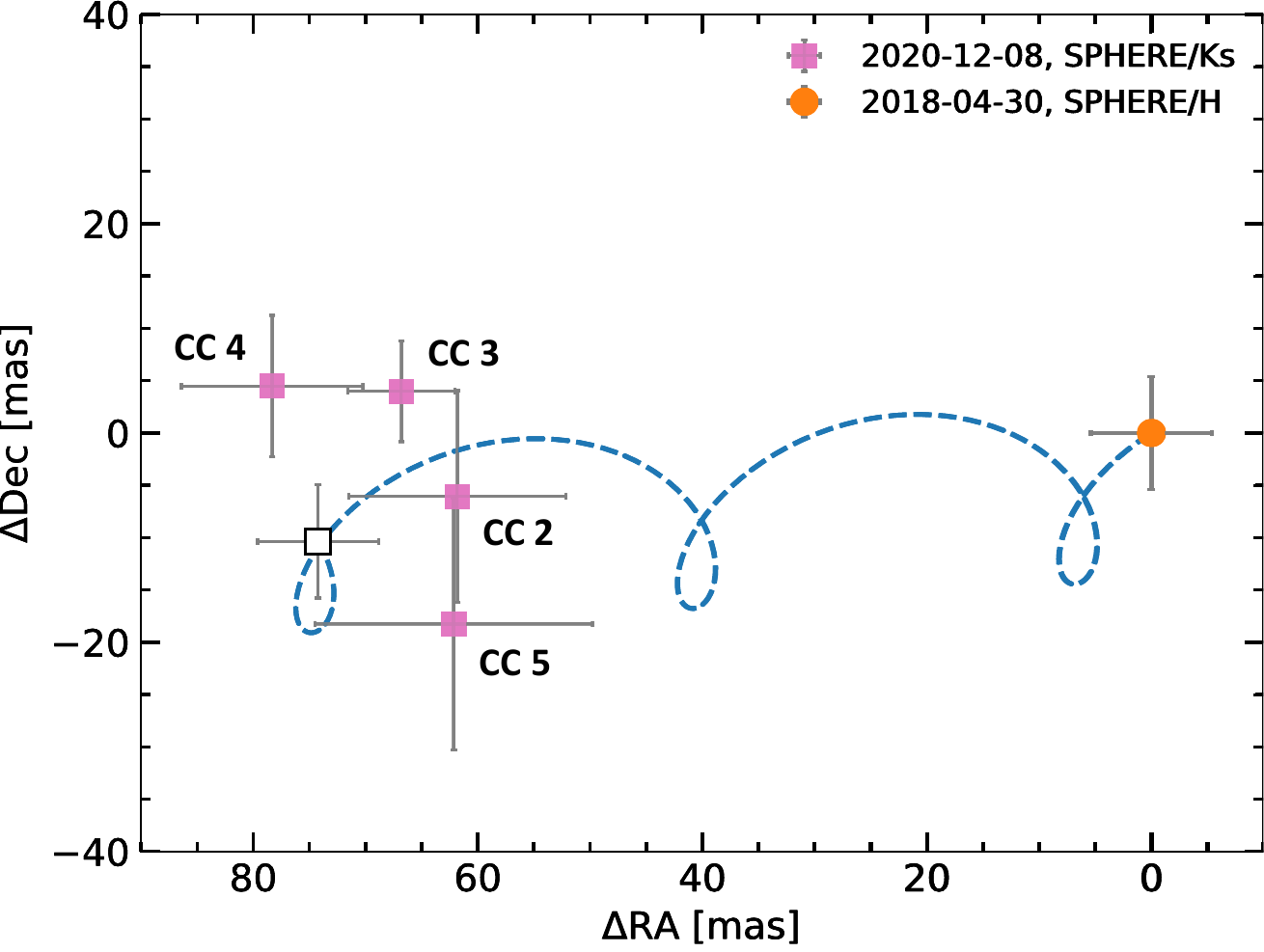}}
\caption{
Proper motion plot for background objects in the SPHERE/IRDIS field of view.
The pink markers indicate the relative astrometric offsets to the first observational epoch that is plotted at the origin of the coordinate system (orange marker).
The blue trajectory represents the simulated motion of a static background object at infinity and the white marker shows the relative positional offset of such an object at the time of our second observation.
}
\label{fig:proper_motion_background}
\end{figure}

In addition to \revrev{YSES~2b}, there are four candidate companions (CCs) in the SPHERE/IRDIS field of view that we could identify in both observational epochs.
These CCs are presented in Fig.~\ref{fig:tyc8984_full}, in which we show the de-rotated data from the night of 2020 December 12.
No PSF subtraction with RDI is performed, instead we just applied an unsharp mask with a Gaussian kernel size of 5 pixels.
\revrev{YSES~2b} can easily be identified in this image product as well.
For the remaining CCs, we present the relative astrometric offsets between both observational epochs in the proper motion diagram in Fig.~\ref{fig:proper_motion_background}.
The relative motions of CCs 2--5 are clearly compatible with stationary background objects, and co-movement can be ruled out for all of them.

\end{appendix}

\end{document}